\documentclass[eqseqnum]{kapproc}%
\usepackage{amsmath}
\usepackage[dvips]{graphicx}%
\usepackage{amsfonts}%
\usepackage{amssymb}
\upperandlowercase 
 \normallatexbib

\begin{document}
\articletitle{Random Matrices and\\ Supersymmetry\\ in Disordered
Systems}
\author{K.B. Efetov}
\affil{Theoretische Physik III, Ruhr-Universität Bochum, 44780
Bochum, Germany,\\ and L.D. Landau Institute for Theoretical
Physics, 117940 Moscow, Russia}
\email{efetov@tp3.rub.de}
\chaptitlerunninghead{Random Matrices and Supersymmetry}

\begin{abstract}
It is described how one comes to the Wigner-Dyson random matrix theory (RMT)
starting from a model of a disordered metal. The lectures start with a
historical introduction where basic ideas of the RMT and theory of disordered
metals are reviewed. This part is followed by an introduction into
supermathematics (mathematics operating with both commuting and anticommuting
variables). The main ideas of the supersymmetry method are given and basic
formulae are derived. Both level-level correlations and fluctuations of
amplitudes of wave functions are discussed. It is shown how one can both
obtain known formulae of the RMT and go beyond. In the last part some recent
progress in the further development of the method and possible perspectives
are discussed.

\end{abstract}


\section*{Introduction}

\subsection{Wigner-Dyson Theory}

According to basic principles of quantum mechanics the energy spectrum of a
particle in a limited volume is discrete. The precise values of the energy
depend on the boundary conditions and the interactions in system. In many
cases these quantities can be calculated with a certain accuracy. However,
often the interactions are so complicated that calculations for the single
levels become impossible. On the other hand, the complexity of the
interactions can lead to the idea of a statistical description in which
information about separate levels is neglected and only averaged quantities
are studied. Density of states, energy level and wave functions correlations
are quantities that can be studied in the statistical approach. Sometimes, it
is sufficient to study the average of, e.g., the density of states or its
variance. In other cases one may be interested in a full statistical
description that can be achieved calculating distribution functions. Studying
the level statistics is, to some extent, analogous to the statistical study of
the motion of atoms and molecules, which is the subject of statistical physics.

The idea of the statistical description of the energy levels was first
proposed by Wigner \cite{wigner} for study of highly excited nuclear levels in
complex nuclei. In such nuclei a large number of particles interact in an
unknown way and the main assumption was that the interactions were equally
probable. Of course, in order to specify the meaning of the words ``equally
probable'' one had to formulate a statistical hypothesis in terms of a
probability distribution that would play the role of the Gibbs distribution.

This was done in Ref. \cite{wigner} in the following way. Choosing a complete
set of eigenfunctions as a basis, one represents the Hamiltonian $H$ as a
matrix with matrix elements $H_{mn}$. The matrix elements $H_{mn}$ are assumed
random with a certain probability distribution. It is clear that the
distribution should not depend of the basis chosen, which implies an invariant
form of the distribution function. In the language of the random matrices
$H_{mn}$ the corresponding distribution function can contain only $Trf\left(
H\right)  $, where $f$ is a function.

The first statistical theory \cite{wigner} was based on a Gaussian
distribution. According to the Gaussian statistical hypothesis a physical
system having $N$ quantum states has the statistical weight $D\left(
H\right)  $%
\begin{equation}
D\left(  H\right)  =A\exp\left[  -\sum_{m,n=1}^{N}\frac{\left|  H_{mn}\right|
^{2}}{2a^{2}}\right]  =A\exp\left[  \frac{-TrH^{2}}{2a^{2}}\right]  \label{a1}%
\end{equation}
In Eq. (\ref{a1}) the parameter $a$ is a cutoff excluding strong interaction,
$H$ is a random $N\times N$ matrix, and $A$ is a normalization coefficient.

It is important to emphasize that the weight $D\left(  H\right)  $ is rather
arbitrary and other forms for the distribution functions can be suggested. Of
course, the dependence of the mean energy level spacing $\Delta\left(
\varepsilon\right)  $
\begin{equation}
\Delta^{-1}\left(  \varepsilon\right)  =\left\langle tr\delta\left(
\varepsilon-H\right)  \right\rangle _{D} \label{a2}%
\end{equation}
on the energy $\varepsilon$ is different for different
distributions $D\left( H\right)  $. (In Eq. (\ref{a2}) the symbol
$\left\langle ...\right\rangle _{D}$ stands for the averaging with
the distribution $D\left(  H\right)  $). For example, the
distribution function $D\left(  H\right)  $ for the Gaussian
distribution, Eq. (\ref{a1}), has the form of a semicircle (Wigner
semicircle law) (for a review, see, e.g. Ref.\cite{mehta}).

What is more interesting, level correlations described, for example, by the
level-level correlation function $R\left(  \omega\right)  $
\begin{equation}
R\left(  \omega\right)  =\Delta^{2}\left(  \varepsilon\right)  \left\langle
Tr\delta\left(  \varepsilon-H\right)  Tr\delta\left(  \varepsilon
-\omega-H\right)  \right\rangle \label{a3}%
\end{equation}
prove to be universal in the limit $N\rightarrow\infty$ provided the energy
$\omega$ is much smaller than the characteristic scale of the variation of
$\Delta\left(  \varepsilon\right)  $. For the Wigner semicircle law, the
latter condition means that the energy $\varepsilon$ is not close to the
points $\varepsilon_{0}$, $-\varepsilon_{0}$, where the quantity $\Delta
^{-1}\left(  \varepsilon\right)  $ proportional to the average density of
states turns to zero.

In the absence of magnetic interactions violating the time reversal symmetry,
the wave functions and matrix elements $H_{mn}$ in Eq. (\ref{a1}) can be
chosen real. In this case the statistical properties of the systems are
described by real symmetric random matrices. The ensemble of real symmetric
matrices with the Gaussian distribution is often called the \textit{Gaussian
orthogonal ensemble }(GOE).

If the magnetic interactions are present in the system, the time reversal and
spin-rotation symmetry is violated and the wave functions are no longer real.
This means that one should deal with general Hermitian matrices without any
additional symmetry and integrate over the matrix elements $H_{mn}$ using only
the constraint $H_{mn}=\left(  H_{nm}\right)  ^{\ast}$. This system is called
the \textit{Gaussian unitary ensemble }(GUE).

The third possible type of the symmetry arises when the system is
time-reversal invariant but does not have central symmetry. In this case it is
also impossible to make all the matrix elements real. Nevertheless an
additional symmetry exists in this case. According to the Kramers theorem all
levels of the system remain doubly degenerate and every eigenvalue of the
matrix $H$ must appear twice. Matrices consisting of real quaternions $H_{mn}$
of the form%
\[
\left(
\begin{array}
[c]{cc}%
p_{mn} & q_{mn}\\
-q_{mn}^{\ast} & p_{mn}^{\ast}%
\end{array}
\right)
\]
and satisfying the condition $H_{mn}=\left(  H_{nm}\right)  ^{+}$ have this
property. The corresponding ensemble is called the \textit{Gaussian symplectic
ensemble }(GSE).

Somewhat different distribution functions were introduced later by Dyson
\cite{dyson} who suggested characterizing the system not by its Hamiltonian
but by an unitary $N\times N$ matrix $S$ whose elements give the transition
probabilities between states, where, again, $N$ is the number of the levels.
This matrix is related to the Hamiltonian $H$ of the system in a complicated
way that is not specified in the theory. According to the Dyson hypothesis the
correlation properties of $n$ successive energy levels of the system $\left(
n\ll N\right)  $ are statistically equivalent to those of $n$ successive
angles provided all the unitary matrices $S$ have equal probabilities. Again,
depending on the symmetry, one can distinguish among the three different
ensembles. The corresponding ensembles are called \textit{Circular ensembles.}

As concerns the complex nuclei, the orthogonal ensembles are most relevant for
their description because in order to change the level statistics one needs,
e.g. huge magnetic fields that hardly exist. However, the other two ensembles
have been under intensive discussions for problems of mesoscopic physics (for
a review, see \cite{ebook}). Moreover, it has been realized that one might
formulate additional ensembles of e.g. chiral matrices relevant for studying
properties of models for QCD (for a review, see the lecture by Jac
Verbaarschot\cite{verbaar}). It has been proven later that in total $10$
different symmetry classes exist \cite{az}. Most of the new ensembles may be
relevant to different disordered mesoscopic systems due the presence of, e.g.,
superconductivity or additional symmetries of the lattice.

I do not plan to discuss in these lectures the non-standard symmetry classes
and restrict myself by the Wigner-Dyson (WD) statistics. It is relevant to say
that calculation of the level-level correlation function, Eq. (\ref{a3}),
starting from the Gaussian or Circular ensembles is not a simple task. The
conventional method of the evaluation is using orthogonal
polynomials\cite{mehta}. The procedure is not difficult for the unitary
ensembles but one has to put a considerable effort to perform the calculations
for the orthogonal and symplectic ones. As it has been mentioned, the final
results are universal in the limit $N\rightarrow\infty$ and can be written for
the orthogonal, unitary and symplectic ensembles in the form
\begin{equation}
R_{orth}\left(  \omega\right)  =1-\frac{\sin^{2}x}{x^{2}}-\frac{d}{dx}\left(
\frac{\sin x}{x}\right)  \int_{1}^{\infty}\frac{\sin xt}{t}dt \label{a4}%
\end{equation}%
\begin{equation}
R_{unit}\left(  \omega\right)  =1-\frac{\sin^{2}x}{x^{2}} \label{a5}%
\end{equation}%
\begin{equation}
R_{sympl}\left(  \omega\right)  =1-\frac{\sin^{2}x}{x^{2}}+\frac{d}{dx}\left(
\frac{\sin x}{x}\right)  \int_{0}^{1}\frac{\sin xt}{t}dt \label{a6}%
\end{equation}
where $x=\pi\omega/\Delta,$ and $\Delta$ is the mean level spacing.

The functions $R\left(  \omega\right)  $, Eqs. (\ref{a4}, \ref{a5}, \ref{a6}),
tend to $1$ in the limit $\omega\rightarrow\infty$, which means that the
correlations are lost in this limit. In the opposite limit $x\rightarrow0$
they turn to zero as $x^{\beta},$ where $\beta=1,2,3$ for the orthogonal,
unitary and symplectic ensembles, respectively. This means that the
probability of finding a level at the distance $\omega$ from another level
decays at small $\omega$. The effect is know as ``level repulsion''. It is
important that Eqs. (\ref{a4}-\ref{a6}) describe the level-level correlation
function, Eq. (\ref{a3}), for both the Gaussian and Circular ensembles.

\subsection{Small disordered particles}

Nowadays the relevance of the random matrix theory (RMT) to mesoscopic physics
is almost evident and it is the starting point of many works on transport in
quantum dots, electromagnetic response of metallic grains, etc. However, it
took quite a long time before the ideas of the RMT penetrated from nuclear to
condensed matter physics.

There were several reasons for this slow development. First, until the end of
60's of the last century most of the objects studied in condensed matter
physics were macroscopic and the discreteness of the energy levels could be
neglected. Therefore the question about the level statistics was not so
interesting. Second, from the theoretical point of view it was not clear at
all how one could come to the Wigner-Dyson level statistics starting from the
Schr\"{o}dinger equation. It was clear that making perturbation theory in
interaction or disorder did not lead to the anything that would resemble Eqs.
(\ref{a4}-\ref{a6}).

As soon as experimentalists started investigation of granular materials (which
happened in 60's), a theory that would describe small metal systems became
nevessary. Such materials consist of small metal particles (grains) with the
diameter down to $10-100$\AA. These grains can be covered by an insulator and
therefore be well isolated from each other. A schematic view of the pattern
can be found in Fig.1
\begin{figure}
[ptb]
\begin{center}
\includegraphics[
height=1.5376in,
width=1.5376in
]%
{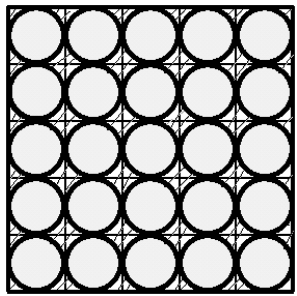}%
\end{center}
\caption[]{
    Structure of a granular metal
}
\end{figure}

It is clear that one can speak now about discrete levels and study their
statistics. Of course, one needs low temperatures in order to prevent
inelastic processes smearing the levels but this is not so difficult
(temperatures $<1K$ can be sufficient).

In practice, the form of the grains can be not very regular, they can contain
defects and impurities and therefore the energy spectrum strongly fluctuates
from grain to grain. All this true even if one neglects the electron-electron
interaction. So, one naturally comes to the idea to describe the levels statistically.

The first work on the application of the RMT to small metallic grains was done
by Gorkov and Eliashberg (GE) \cite{ge}. These authors studied the
electromagnetic response of the system of the grains and therefore they needed
an information about the level-level correlation in a single grain. As the RMT
is purely phenomenological and nothing is assumed about the origin of the
randomness, this theory was taken by GE to describe the correlations. Starting
from the explicit form of the level-level correlation function $R\left(
\omega\right)  $, Eqs. (\ref{a4}-\ref{a6}), they calculated the desired
physical quantity.

GE identified correctly physical situations when the three symmetry classes
might be used. In the absence of any magnetic and spin orbit interactions they
suggested to use the orthogonal ensemble. If a magnetic field is applied or
there are magnetic impurities in the grains, the unitary ensemble should be
applicable. If there are no magnetic interactions but spin-orbital impurities
are present, the grains should be described by the symplectic ensemble.

Being the first application of the RMT in solid state physics, the
paper\cite{ge} remained the only application during the next $17$ years. This
is not surprising because using calculational schemes existed in that time no
indication in non-trivial level correlations could be seen. Let us discuss
this point in more details.

Studying a disordered system and neglecting electron-electron interactions one
can start with the following Hamiltonian%
\begin{equation}
H=\varepsilon\left(  \mathbf{\hat{p}}\right)  +U\left(  \mathbf{r}\right)
\label{a8}%
\end{equation}
where $\varepsilon\left(  \mathbf{\hat{p}}\right)  $ is the operator of the
kinetic energy and calculate the Green function $G_{\varepsilon}=\left(
\varepsilon-H\right)  ^{-1}$ performing an expansion in the disorder potential
$U\left(  \mathbf{r}\right)  $. Usually, it is assumed that the $U\left(
\mathbf{r}\right)  $ is random and its fluctuations are Gaussian with
\begin{equation}
\left\langle U\left(  \mathbf{r}\right)  \right\rangle =0\text{, }\left\langle
U\left(  \mathbf{r}\right)  U\left(  \mathbf{r}^{\prime}\right)  \right\rangle
=\frac{1}{2\pi\nu\tau}\delta\left(  \mathbf{r-r}^{\prime}\right)  \label{a9}%
\end{equation}
where $\nu$ is the density of states and $\tau$ is the mean scattering time.

Making the perturbation theory in the random potential $U\left(
\mathbf{r}\right)  $ and averaging over this potential can be done using the
``cross technique'' \cite{agd}. Only diagrams without intersections of the
impurity lines are important in the standard approximation of the weak
disorder. For the one particle Green function the typical diagrams are
represented in Fig.2.%
\begin{figure}
[ptb]
\begin{center}
\includegraphics[
height=1.7694in,
width=2.3575in
]%
{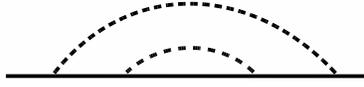}%
\end{center}
\caption[]{
  Typical diagrams for the Green function in a disordered metal
}
\end{figure}

As a result, one obtains for the averaged Green function $\left\langle
G\right\rangle $ the following expression%
\begin{equation}
G=\frac{1}{\varepsilon-\varepsilon\left(  \mathbf{p}\right)  \pm i/2\tau}
\label{a10}%
\end{equation}
where ``$+"$ corresponds to the retarded and ``$-"$ to the advanced Green functions.

Standard calculations based on a summation of ``ladder diagrams'' lead in this
case to the classical Drude formula for, e.g., the conductivity $\sigma\left(
\omega\right)  $
\begin{equation}
\sigma\left(  \omega\right)  =\frac{\sigma_{0}}{1-i\omega\tau},\text{ }%
\sigma_{0}=2e^{2}\nu D_{0} \label{a7}%
\end{equation}
where $D_{0}=v_{0}^{2}\tau/3$ is the classical diffusion coefficient, $\omega$
is the frequency and $e$ is the electron charge.

So long as the grain size remained larger than the atomic distances, no
deviations from this formula could be found and therefore the question about
the applicability of the RMT remained open. It is relevant to mention works by
Kubo performed in approximately the same time. In Ref. \cite{kubo1}he argued
that even very small irregularities (with size of the order of atomic
distances) of the shape of the metallic grains must lead to lifting of all
degeneracies of eigenstates that are present in ideally spherical particles.
This lead him to the conclusion that the mean energy level spacing $\Delta$ is
inversely proportional to the volume $V$ of the particle
\begin{equation}
\Delta=\left(  \nu V\right)  ^{-1} \label{b1}%
\end{equation}
where $\nu$ is density of states at the Fermi surface of the metal. Later Kubo
suggested \cite{kubo2} that the spacing distribution had to follow the Poisson
law. The latter differs essentially from the Wigner-Dyson RMT by the absence
of the level repulsion. The second work by Kubo, Ref.\cite{kubo2}, was
published several years after the work by Gorkov and Eliashberg, Ref.
\cite{ge}, and this shows that the applicability of the Wigner-Dyson
statistics to the small metal particles was far from being established in that time.

The situation started to change only at the end of 70's with the new
developments in the theory of Anderson localization. In the
publication\cite{aalr} a new scaling idea was put forward for description of
disordered samples of an arbitrary dimensionality. The most unusual was a
prediction that two dimensional initially metallic samples could not remain
metals in the presence of an arbitrary weak disorder and had to acquire
insulating properties. Again, using Eq. (\ref{a7}) it was not clear why
something had to happen in two dimension (the localization in one-dimensional
chains had been proven before and it was clear that the ladder diagrams
summation leading to Eq. (\ref{a7}) was not sufficient for that case).

Trying to understand how something unusual could happen in two dimensions
Gorkov, Larkin and Khmelnitskii \cite{glk} investigated more complicated
diagrams and found that a certain class of diagrams could lead to a divergence
in any dimensionality $d\leqslant2$. These are ``fan'' diagrams with the
maximal number of crossings. They are represented in Fig. 3
\begin{figure}
[ptb]
\begin{center}
\includegraphics[
height=1.7642in,
width=2.3506in
]%
{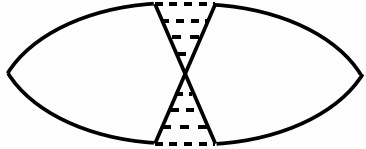}%
\end{center}
\caption[]{Cooperon: a singular correction to conductivity

}
\end{figure}
and their sum is a new effective mode that is usually called ``cooperon''.
This mode has a form of a diffusion propagator and its contribution to the
conductivity can be written in the form%
\begin{equation}
\sigma\left(  \omega\right)  =\sigma_{0}\left(  1-\frac{1}{\pi\nu}\int\frac
{1}{D_{0}\mathbf{k}^{2}-i\omega}\frac{d^{d}\mathbf{k}}{\left(  2\pi\right)
^{d}}\right)  \label{a11}%
\end{equation}
It is clear from Eq. (\ref{a11}) that in the limit of low frequencies
$\omega\rightarrow0$ the second term in the brackets in Eq. (\ref{a11})
diverges in any dimension $d\leqslant2.$ This means that in a disordered film
the quantum correction to the classical conductivity diverges and this signals
(but, of course, does not prove) the possibility of the localization.

The scaling theory of the localization and the discovery of the new diffusion
modes was revolutionary in the theory of disordered systems but how can these
findings be related to the Wigner-Dyson theory?

Actually, Eq. (\ref{a11}) is the key to understanding that there can be
something beyond the classical Eq. (\ref{a7}) in small metal particles. One
needs only realizing that the case of small metal grains corresponds to the
zero dimensionality of the integral in Eq. (\ref{a11}). Due to the finite size
the values of the momentum $\mathbf{k}$ are quantized such that $k_{a}=2\pi
n_{a}/L_{a}$, where $n_{a}=0,$ $\pm1,$ $\pm2,$ $\pm3......$ , $a=\left(
x,y,z\right)  $, and $L_{a}$ is the size of the grain in the $a$ -direction.

At low frequencies $\omega\ll D_{0}/L^{2}$ the most important contribution in
the integral in Eq. (\ref{a11}) comes from the zero harmonics with
$\mathbf{k}=0$ and one can see that this contribution strongly diverges when
$\omega\rightarrow0$. Moreover, the quantum correction is proportional to
$\Delta/\omega$, where $\Delta$ is given by Eq. (\ref{b1}) and this is what
one can expect from Eqs. (\ref{a4}-\ref{a6}).

So, Eq. (\ref{a11}) really signals that something nontrivial can happen in
metal grains and the WD theory is not excluded. The diffusion modes play a
prominent role and it seems, at first glance, that one should merely write
proper diagrams and sum their contribution. However, even if this were
possible this would hardly correspond to Eq. (\ref{a4}-\ref{a6}). The problem
is that the expansion in terms of $\Delta/\omega$ cannot take into account the
oscillating part in Eqs. (\ref{a4}-\ref{a6}) even in principle. Summing the
diagrams one can hope to reproduce only non-oscillating asymptotics of these
equations. This means that another approach has to be developed.

The possibility to demonstrate that the energy level and wave functions
statistics can really be described by the RMT came first with the development
of the supersymmetry approach \cite{efetov1,ereview}. This method is based on
a representation of Green functions of a disordered metal in terms of an
integral over both commuting and anticommuting variables. Singling out
excitations with the lowest energy (diffusion modes) one can reduce
calculations to a supermatrix non-linear model. The supersymmetry method
allowed to prove for the first time that the level-level correlation function
for disordered particles is really described by Eqs. (\ref{a4}-\ref{a6}).
Later, using the supersymmetry technique Verbaarschot, Weidenm\"{u}ller and
Zirnbauer \cite{vwz} have derived Eqs. (\ref{a4}-\ref{a6}) starting directly
from the Gaussian ensembles, Eq. (\ref{a1}).

When deriving the non-linear $\sigma$-model for metallic particles it was very
important that they contained disorder. However, it is not the necessary
condition. Several years later Bohigas, Gianonni and Schmidt \cite{bgs}
conjectured that the RMT should describe correctly spectral properties of
quantum systems which are chaotic in their classical limit. In particular, the
Wigner-Dyson statistics had to be observed in clean metal particles (quantum
billiards) provided their shape was such that classical motion would be
chaotic. Their hypothesis was made on the basis of extensive numerics. For a
review of the subsequent activity in these fields the book \cite{suptrace} is
a good reference.

Historically, the description of disordered systems with a non-linear $\sigma
$-model has been suggested by Wegner \cite{wegner} using the replica method
and integration over conventional complex variables. In the first work there
were problems with convergence of functional integrals and therefore the
replica approach was further developed in the publications \cite{elk} and
\cite{sw}. The $\sigma$-model of Ref. \cite{sw} was obtained by the
integration over conventional variables and, as a result, the group of the
matrices $Q$ was non-compact. In contrast, the starting point of Ref.
\cite{elk} was a representation of Green functions in terms of integrals over
anticommuting (Grassmann) variables and this lead to a compact group of the
matrices $Q$.

Although both the replica and supersymmetry approach are equivalent when doing
the perturbation theory in the diffusion modes, the latter method is much more
efficient for non-perturbative calculations like the study of the level-level
correlations. This had become clear shortly after the works \cite{elk,sw} were
finished and this drawback of the replica approach motivated the development
of the supersymmetry one. It should be noticed that recently the oscillating
behavior of the level-level correlation function $R\left(  \omega\right)  $
has been obtained \cite{km,yl,kanzieper} using the compact replica $\sigma
$-model of Ref. \cite{elk}. However, the procedures used in these references
are considerably more complicated than the calculations by the supersymmetry
method and the limit of low frequencies $\omega\lesssim\Delta$ is still hardly achievable.

It is fair to say that the replica approach allows including electron-electron
interactions in a comparatively easy way \cite{finkelstein,fin1} and this has
been the main motivation in the attempts \cite{km,yl,kanzieper} to obtain
non-perturbative results within the replica technique. At the same time, it
was believed for a long time that an inclusion of the electron-electron
interaction into the supersymmetry scheme was impossible. However, this is not
quite so and, at least, not very strong interaction can be incorporated in the
supermatrix $\sigma$-model \cite{schwiete}.

It follows from this discussion that the supersymmetry approach is better
suitable for making connections with the RMT and therefore the present
lectures contain discussions based on this method only. It is possible neither
review here all works made in this direction nor present all details of the
calculations. For a more detailed information see the book \cite{ebook} and
more recent reviews \cite{fyodorov,gmw,mirlin,ast,asz}. It is relevant to
mention here that the word \textit{supersymmetry }has appeared in the
condensed matter physics in the publication by Parisi and Sourlas \cite{ps},
who discovered a complex symmetry in a model describing ferromagnets in a
random magnetic field. They used a concept of superspace including both
commuting and anticommuting variables.

Two other related directions of the use of the RMT are reviewed at this school
by Boris Altshuler (Quantum Chaos) and Jac Verbaarschot (QCD).

In the next sections I want to present the main ideas of the supersymmetry
approach and show how it can be used for both the level correlations and wave
functions statistics. It will be shown how to obtain the Wigner-Dyson
statistics and how to go beyond it. A new development concerning a
generalization of the supersymmetric $\sigma$-model to more complicated
situations will be outlined in the last section.

\section{Supersymmetry method.}

\subsection{Supermathematics}

The supersymmetry method is based on the use of the so called Grassmann
variables $\chi_{i}$, $i=1,$ $2,...,$ $n$ (the elements of the Grassmann
algebra) that are introduced in a completely formal way. These are abstract
objects but in many cases abstract mathematical constructions drastically
influence the development of physics. For example, nobody can dispute the
usefulness of complex numbers for physics, but what is the physical meaning of
$\sqrt{-1}$? Here I want to remind the reader basic formulae concerning
definitions and operations with objects containing combinations of the
Grassmann variables and conventional numbers (supermathematics).

The Grassmann variables are some mathematical objects obeying the following
anticommutation rules \cite{berezin}
\begin{equation}
\left\{  \chi_{i},\chi_{j}\right\}  =\chi_{i}\chi_{j}+\chi_{j}\chi_{i}=0
\label{c1}%
\end{equation}
for any $1\leq i,$ $j\leq n.$

The anticommutation rules, Eq. (\ref{c1}) hold in particular for $i=j$ and we
see that the square of an arbitrary variable $\chi_{i}$ is zero%
\begin{equation}
\chi_{i}^{2}=0 \label{c2}%
\end{equation}
For any anticommuting variable $\chi$ one can introduce its ``complex
conjugate) $\chi^{\ast}$. It is assumed by the definition that $\left(
\chi^{\ast}\right)  ^{\ast}=-\chi$, such that the ``square of the modulus is
``real''%
\begin{equation}
\left(  \chi_{i}^{\ast}\chi_{i}\right)  ^{\ast}=-\chi_{i}\chi_{i}^{\ast}%
=\chi_{i}^{\ast}\chi_{i} \label{c3}%
\end{equation}

The anticommuting variables $\chi_{i}$, Eq. (\ref{c1}-\ref{c3}), remained not
very useful until Berezin introduced integrals over these variables. The
integrals are nothing more than formal symbols introduced as follows
\begin{equation}
\int d\chi_{i}=\int d\chi_{i}^{\ast}=0,\text{ }\int\chi_{i}d\chi_{i}=\int
\chi_{i}^{\ast}d\chi_{i}^{\ast}=1 \label{c4}%
\end{equation}
It is implied that the ``differentials'' $d\chi_{i}$, $d\chi_{i}^{\ast}$
anticommute with each other and with the variables $\chi_{i},$ $\chi_{i}%
^{\ast}$%
\begin{align}
\left\{  d\chi_{i},d\chi_{j}\right\}   &  =\left\{  d\chi_{i},d\chi_{j}^{\ast
}\right\}  =\left\{  d\chi_{i}^{\ast},d\chi_{j}^{\ast}\right\}  =0\label{c5}\\
\left\{  d\chi_{i},\chi_{j}\right\}   &  =\left\{  d\chi_{i},\chi_{j}^{\ast
}\right\}  =\left\{  d\chi_{i}^{\ast},\chi_{j}\right\}  =\left\{  d\chi
_{i}^{\ast},\chi_{j}^{\ast}\right\}  =0\nonumber
\end{align}
The definition, Eq. (\ref{c4}), is sufficient for introducing integrals of an
arbitrary function. If such a functions depends only on one variable $\chi
_{i}$ it must be linear in $\chi_{i}$ because already $\chi_{i}^{2}=0.$
Assuming that the integral of a sum of two functions equals the sum of the
integrals we calculate the integral of the sum with Eq. (\ref{c4}). The
repeated integrals are implied by integrals over several variables. This
enables us to calculate the integral of a function of an arbitrary number of variables.

The most important for the development of the supersymmetry method are
Gaussian integrals. The direct integration shows that the following relation
is fulfilled%
\begin{equation}
I=\int\exp\left(  -\chi^{+}A\chi\right)  \prod_{i=1}^{n}d\chi_{i}^{\ast}%
d\chi_{i}=DetA \label{c6}%
\end{equation}
where $A$ is an $n\times n$ Hermitian matrix and%
\begin{equation}
\chi=\left(
\begin{array}
[c]{c}%
\chi_{1}\\
\chi_{2}\\
.\\
.\\
\chi_{n}%
\end{array}
\right)  \text{, }\chi^{+}=\left(
\begin{array}
[c]{ccccc}%
\chi_{1}^{\ast} & \chi_{2}^{\ast} & . & . & \chi_{n}^{\ast}%
\end{array}
\right)  \label{c7}%
\end{equation}
Eq. (\ref{c6}) differs from the corresponding equation for the commuting
variables by giving $\det A$ instead $\left(  \det A\right)  ^{-1}.$ This
remarkable difference is the basis of the supersymmetry method presented in
these lectures. In addition to Eq. (\ref{c6}) one can write one more useful
integral%
\begin{equation}
I_{2}=\frac{\int\chi_{i}\chi_{k}^{\ast}\exp\left(  -\chi^{+}A\chi\right)
\prod_{l=1}^{n}d\chi_{l}^{\ast}d\chi_{l}}{\int\exp\left(  -\chi^{+}%
A\chi\right)  \prod_{l=1}^{n}d\chi_{l}^{\ast}d\chi_{l}}=\left(  A^{-1}\right)
_{ik} \label{c8}%
\end{equation}
In contrast to the integral $I,$ Eq. (\ref{c6}), the integral $I_{2},$ Eq.
(\ref{c8}), is completely similar to the corresponding integral over
conventional numbers. Eq. (\ref{c8}) can be proven by the differentiation of
$\ln I$ in $A_{ki}.$

The next step is the introduction of supervectors and supermatrices. An $n+m$
component supervector is introduced as
\begin{equation}
\Phi=\left(
\begin{array}
[c]{c}%
\chi\\
S
\end{array}
\right)  \label{c9}%
\end{equation}
where the $n$- component vector $\chi$ is defined in Eq. (\ref{c7}). The
$m$-component vector $S$ has a similar form%
\begin{equation}
\left(
\begin{array}
[c]{c}%
S_{1}\\
S_{2}\\
.\\
.\\
S_{m}%
\end{array}
\right)  \label{c10}%
\end{equation}
but its components are conventional complex numbers.

In analogy with conventional vectors one can introduce the Hermitian
conjugation
\begin{equation}
\Phi^{+}=\left(  \Phi^{T}\right)  ^{\ast} \label{c11}%
\end{equation}
and the scalar product%
\begin{equation}
\Phi^{i+}\Phi^{j}=\sum_{\alpha=1}^{n}\chi_{\alpha}^{i\ast}\chi_{\alpha}%
^{j}+\sum_{\alpha=1}^{m}S_{\alpha}^{i\ast}S_{\alpha}^{j} \label{c12}%
\end{equation}
A linear transformation $F$ in the space of the supervectors converts a
supervector $\Phi$ into another supervector $\tilde{\Phi}$%
\begin{equation}
\tilde{\Phi}=F\Phi\label{c13}%
\end{equation}
Of course, the supervector $\tilde{\Phi}$ must have the same structure, Eq.
(\ref{c9}), as the supervector $\Phi.$ This imposes a restriction on the
structure of the supermatrix $F$ corresponding to the linear transformation
$F:$ it has to be of the form%
\begin{equation}
F=\left(
\begin{array}
[c]{cc}%
a & \sigma\\
\rho & b
\end{array}
\right)  \label{c14}%
\end{equation}
In Eq. (\ref{c14}) $a$ and $b$ are $n\times n$ and $m\times m$ matrices
containing only commuting variables, $\sigma$ and $\rho$ are $n\times m$ and
$m\times n$ matrices consisting of anticommuting ones. Matrices having the
structure, Eq. (\ref{c14}) can be called supermatrices.

Two supermatrices $F$ and $G$ of the rank $\left(  m+n\right)  \times\left(
n+m\right)  $ are assumed to multiply according to the conventional rules%
\begin{equation}
\left(  FG\right)  _{ik}=\sum_{l=1}^{m+n}F_{il}G_{lk} \label{c15}%
\end{equation}
and one can see that $FG$ is a supermatrix of the same form. In order to
define the supertranspose $F^{T}$ of the supermatrix $F$ one should use the
notion of the scalar product of two supervectors, Eq. (\ref{c12}). Again, by
analogy with the conventional definition the supermatrix $F^{T}$ is introduced
as%
\begin{equation}
\Phi_{1}^{T}F^{T}\Phi_{2}=\left(  F\Phi_{1}\right)  ^{T}\Phi_{2} \label{c16}%
\end{equation}
The transpose of a conventional matrix is obtained by transposing its indexes.
This is not as simple for the supermatrices. Writing out the scalar product on
both sides of Eq. (\ref{c16}) explicitly and using the anticommutation
relation, Eq. (\ref{c1}), one can see that the supermatrix $F^{T}$ is equal to%
\begin{equation}
F^{T}=\left(
\begin{array}
[c]{cc}%
a^{T} & -\rho^{T}\\
\sigma^{T} & b^{T}%
\end{array}
\right)  \label{c17}%
\end{equation}
where $a^{T},$ $b^{T},$ $\sigma^{T},$ and $\rho^{T}$ stand for the
conventional transposition of the matrices $a,$ $b,$ $\sigma,$ and $\rho$.

Using the scalar product, Eq. (\ref{c16}) one obtains immediately%
\begin{equation}
\left(  F_{1}F_{2}\right)  ^{T}=F_{2}^{T}F_{1}^{T} \label{c18}%
\end{equation}
The Hermitian conjugate $F^{+}$ of the matrix $F$ can be defined in a standard
way%
\begin{equation}
F^{+}=\left(  F^{T}\right)  ^{\ast} \label{c19}%
\end{equation}
Combining Eqs. (\ref{c3}, \ref{c18}, \ref{c19}) one can obtain%
\begin{equation}
\left(  F_{1}F_{2}\right)  ^{+}=F_{2}^{+}F_{1}^{+}\text{ and }\left(
F^{+}\right)  ^{+}=F \label{c20}%
\end{equation}
The latter equality shows that the operation of the Hermitian conjugation is
inverse to itself. The same is not generally true for the transposition%
\begin{equation}
\left(  F^{T}\right)  ^{T}\neq F \label{c21}%
\end{equation}
A very important operation in the theory of conventional matrices is taking
the trace of a matrix. If one takes the trace of a product of several matrices
it is invariant under cyclic permutations of the matrices. However, due to the
presence of anticommuting elements a proper operation for the supermatrices
should be defined in a different way. The supertrace $STrF$ of matrix of the
form, Eq.(\ref{c14}) is defined as
\begin{equation}
STrF=Tra-Trb \label{c22}%
\end{equation}
where the symbol $Tr$ stands for the conventional trace.

Although somewhat strange, the definition, Eq. (\ref{c22}) is very useful
because it is this operation that provides the invariance under the cyclic
permutations. We obtain for arbitrary supermatrices $F_{i}$ of the form, Eq.
(\ref{c14})%
\begin{equation}
STrF_{1}F_{2}=STrF_{2}F_{1} \label{c23}%
\end{equation}
and%
\begin{equation}
STr\left(  F_{1}F_{2}...F_{n}\right)  =STr\left(  F_{n}F_{1}F_{2}%
...F_{n-1}\right)  \label{c24}%
\end{equation}
In addition to the supertrace it is convenient to introduce a superdeterminant
of the supermatrix $F$%
\begin{equation}
\ln SDetF=STr\ln F \label{c25}%
\end{equation}
The superdeterminant $SDetF$ can also be written as%
\begin{equation}
SDetF=Det\left(  a-\sigma b^{-1}\rho\right)  Detb^{-1} \label{c26}%
\end{equation}
The connection between the superdeterminant and supertrace enables us to prove
immediately the multipliticity of the superdeterminant%
\begin{equation}
SDet\left(  F_{1}F_{2}\right)  =\left(  SDetF_{1}\right)  \left(
SDetF_{2}\right)  \label{c27}%
\end{equation}
The rules of the operations with the supervectors and supermatrices are very
convenient because they are similar to those of conventional linear algebra.
In fact, one can manipulate superobjects in exactly the same way as
conventional objects. This simplifies calculations with quantities containing
both types of variables considerably.

Now we can write Gaussian integrals over supervectors that generalize the
integrals over conventional complex numbers or Grassmann variable. A direct
calculation shows that%
\begin{equation}
I^{s}=\int\exp\left(  -\Phi^{+}F\Phi\right)  d\Phi^{\ast}d\Phi=SDetF,
\label{c28}%
\end{equation}%

\[
d\Phi^{\ast}d\Phi=\pi^{-m}\prod_{i=1}^{n}d\chi_{i}^{\ast}d\chi_{i}\prod
_{k=1}^{m}dS_{i}^{\ast}dS_{i}%
\]
and
\begin{equation}
I_{2}^{s}=\frac{\int\Phi_{i}\Phi_{k}^{\ast}\exp\left(  -\Phi^{+}F\Phi\right)
d\Phi^{\ast}d\Phi}{\int\exp\left(  -\Phi^{+}F\Phi\right)  d\Phi^{\ast}d\Phi
}=\left(  F^{-1}\right)  _{ik} \label{c29}%
\end{equation}

The formulae written in this subsection give complete information about
integrals over the Grassmann variables, supervectors, and supermatrices. This
information will be directly used for constructing the supersymmetry method.

\subsection{Physical quantities as integrals over supervectors. Averaging over disorder.}

Eq. (\ref{c29}) enables us to express physical quantities in terms of
functional integrals over supervectors. The form of the integrals that will be
obtained is such that averaging over disorder can be performed the beginning
of all calculations.

I start with the Schr\"{o}dinger equation for electrons without any
electron-electron interactions but in a presence of an external potential
containing both regular and irregular parts. The regular part can describe
potential walls and other features of the system whereas the irregular part
$H_{1}$ of the Hamiltonian stands for disorder. The Schr\"{o}dinger equation
takes the form%
\begin{equation}
H\phi_{k}=\varepsilon_{k}\phi_{k},\text{ }H=H_{0}+H_{1}\text{, }\left\langle
H_{1}\right\rangle =0 \label{c30}%
\end{equation}
where $\phi_{k}$ and $\varepsilon_{k}$ are eigenfunctions and eigenvalues,
respectively. The angular brackets $\left\langle ...\right\rangle $ stand for
the averaging over disorder.

The most important physical quantities can be expressed in terms of retarded
$G_{\varepsilon}^{R}$ and advanced $G_{\varepsilon}^{A}$ Green functions of
the Schr\"{o}dinger equation. Using the spectral expansion the Green functions
$G_{\varepsilon}^{R,A}$ can be written in the form
\begin{equation}
G_{\varepsilon}^{R,A}\left(  \mathbf{r,r}^{\prime}\right)  =\sum_{k}\frac
{\phi_{k}\left(  \mathbf{r}\right)  \phi_{k}^{\ast}\left(  \mathbf{r}^{\prime
}\right)  }{\varepsilon-\varepsilon_{k}\pm i\delta}=\sum_{k}G_{\varepsilon
k}^{R,A}\phi_{k}\left(  \mathbf{r}\right)  \phi_{k}^{\ast}\left(
\mathbf{r}^{\prime}\right)  \label{c31}%
\end{equation}
These functions satisfy the equation%
\begin{equation}
\left(  \varepsilon-H\right)  G_{\varepsilon}^{R,A}\left(  \mathbf{r,r}%
^{\prime}\right)  =\delta\left(  \mathbf{r-r}^{\prime}\right)  \label{c31a}%
\end{equation}

The average density of states $\rho\left(  \varepsilon\right)  $ (this
quantity is proportional to $\Delta^{-1}$, Eq. (\ref{a2})) can be written as%
\begin{align}
\left\langle \rho\left(  \varepsilon,\mathbf{r}\right)  \right\rangle  &
=\left\langle \sum_{k}\phi_{k}^{\ast}\left(  \mathbf{r}\right)  \phi
_{k}\left(  \mathbf{r}\right)  \delta\left(  \varepsilon-\varepsilon
_{k}\right)  \right\rangle \label{c32}\\
&  =\frac{1}{\pi}\left\langle \operatorname{Im}G_{\varepsilon}^{A}\left(
\mathbf{r,r}\right)  \right\rangle
\end{align}
whereas the level-level correlation function $R\left(  \omega\right)  $ takes
the form%
\begin{equation}
R\left(  \omega\right)  =\left(  \frac{\Delta}{\pi}\right)  ^{2}\left\langle
\sum_{k,m}\operatorname{Im}G_{k,\varepsilon-\omega}^{A}\operatorname{Im}%
G_{m,\varepsilon}^{A}\right\rangle \label{c33}%
\end{equation}

We see from Eqs. (\ref{c32}, \ref{c33}) that, as soon as we are able to
average the Green functions or their products over the disorder, the
quantities of interest are found. However, this cannot be done directly using
Eqs. (\ref{c32}, \ref{c33}) and we need another representation for the Green
functions. Of course, one can do perturbation theory in the disorder potential
but, as it has been already discussed, such an approach can hardly help in
obtaining the Wigner-Dyson statistics.

What will be done now is writing the Green functions in a form that would be
suitable for averaging over the disorder in the very beginning. This can be
conveniently done with the integrals over the supervectors. I want to present
here the main scheme only. All necessary details can be found in the book
\cite{ebook}. As the main interest is to calculate the level-level correlation
function $R\left(  \omega\right)  ,$ Eq. (\ref{c33}), all formulae will be
written for this case. Calculating the density of states, Eq. (\ref{c32}), is
a simpler and less interesting task.

Let us introduce $8$-component supervectors $\psi$ consisting of $4$-component
supervectors $\psi^{1}$ and $\psi^{2}$ such that
\begin{equation}
\psi^{m}=\left(
\begin{array}
[c]{c}%
\theta^{m}\\
v^{m}%
\end{array}
\right)  ,\text{ }\theta^{m}=\frac{1}{\sqrt{2}}\left(
\begin{array}
[c]{c}%
\chi^{m\ast}\\
\chi^{m}%
\end{array}
\right)  ,\text{ }v^{m}=\frac{1}{\sqrt{2}}\left(
\begin{array}
[c]{c}%
S^{m\ast}\\
S^{m}%
\end{array}
\right)  , \label{c34}%
\end{equation}
$m=1,2.$

For the supervectors $\psi$ of the form of Eq. (\ref{c34}) one can define, in
addition to transposition and Hermitian conjugation, the operation of the
``charge conjugation''%
\begin{equation}
\bar{\psi}=\left(  C\psi\right)  ^{T},\text{ }\bar{\psi}^{m}=\left(
\begin{array}
[c]{cc}%
\bar{\theta}^{m}, & v^{m}%
\end{array}
\right)  \label{c35}%
\end{equation}
In Eq. (\ref{c35}), $T$ stands for transposition, and $C$ is the supermatrix
of the form%
\[
C^{mn}=\Lambda^{mn}\left(
\begin{array}
[c]{cc}%
c_{1} & 0\\
0 & c_{2}%
\end{array}
\right)
\]
where $\Lambda$ is the diagonal supermatrix%
\begin{equation}
\Lambda=\left(
\begin{array}
[c]{cc}%
\mathbf{1} & 0\\
0 & -\mathbf{1}%
\end{array}
\right)  \label{c36}%
\end{equation}
with $\mathbf{1}$ the unity $4\times4$ unity matrix.

The matrices $c_{1}$ and $c_{2}$ have the form%
\begin{equation}
c_{1}=\left(
\begin{array}
[c]{cc}%
0 & -1\\
1 & 0
\end{array}
\right)  ,\text{ }c_{2}=\left(
\begin{array}
[c]{cc}%
0 & 1\\
1 & 0
\end{array}
\right)  \label{c37}%
\end{equation}
The function $R\left(  \omega\right)  $ is determined by products $G^{A}G^{R},
$ $G^{R}G^{R}$ and $G^{A}G^{A}.$ The last two products are not interesting
because the average of their products is equal for weak disorder to the
product of their average. When calculating these quantities the diffusion
modes discussed in the Introduction do not appear. Therefore let us
concentrate on the calculation of the product $G^{A}G^{R}.$ Using Eqs.
\ref{c28}, \ref{c29}) we can write this quantity as%
\begin{equation}
G_{\varepsilon-\omega}^{A}\left(  \mathbf{r,r}\right)  G_{\varepsilon}%
^{R}\left(  \mathbf{r}^{\prime},\mathbf{r}^{\prime}\right)  =\left(
\frac{\Delta}{\pi}\right)  ^{2}\int\psi^{1}\left(  \mathbf{r}\right)
\bar{\psi}^{1}\left(  \mathbf{r}\right)  \psi^{2}\left(  \mathbf{r}^{\prime
}\right)  \bar{\psi}^{2}\left(  \mathbf{r}^{\prime}\right)  \exp\left(
-L\right)  D\psi\label{c38}%
\end{equation}
where the Lagrangian $L$ has the form%
\begin{equation}
L=i\int\bar{\psi}\left(  \mathbf{r}\right)  \left(  -\tilde{H}_{0}-U\left(
\mathbf{r}\right)  -\frac{1}{2}\left(  \omega+i\delta\right)  \Lambda\right)
\psi\left(  \mathbf{r}\right)  d\mathbf{r} \label{c39}%
\end{equation}
where $U\left(  \mathbf{r}\right)  $ the impurity potential. The operator
$\tilde{H}_{0}$ equals%
\begin{equation}
\tilde{H}_{0}=\varepsilon\left(  -i\nabla_{r}\right)  -\varepsilon
+\frac{\omega}{2} \label{c40}%
\end{equation}
where $\varepsilon\left(  -i\nabla_{r}\right)  $ is the spectrum.

I would like to draw attention at this point that the weight denominator is
absent in Eq. (\ref{c38}), which contrasts the analogous integrals, Eqs.
(\ref{c8},\ref{c29}). Actually, this is a consequence of the fact that $F$ in
Eq. (\ref{c29}) is taken as unity in the space of the $2\times2$ supermatrices
when writing Eq. (\ref{c38}). In other words, the weight denominator is absent
due to the difference of the results of the Gaussian integration over the
anticommuting variables $\chi,$ Eq. (\ref{c6}), (one obtains $DetA$) and the
corresponding formula for the integration over conventional numbers that gives
$\left(  DetA\right)  ^{-1}$.

The possibility to write the Green function in the form of Eq. (\ref{c38})
without the weight denominator is the reason why the integration over the
supervectors is used. As the weight denominator is absent one can immediately
average over impurities in Eq. (\ref{c38}).

Let us assume for simplicity that the distribution of the random potential
$U\left(  \mathbf{r}\right)  $ in Eq. (\ref{c39}) is specified by Eq.
(\ref{a9})$.$ Then, the averaging over the random potential is simple and one
obtains again Eq. (\ref{c38}) but now the Lagrangian $L$ should be written as%
\begin{equation}
L=\int\left[  -i\bar{\psi}\tilde{H}_{0}\psi+\frac{1}{4\pi\nu\tau}\left(
\bar{\psi}\psi\right)  ^{2}-\frac{i\left(  \omega+i\delta\right)  }{2}%
\bar{\psi}\Lambda\psi\right]  d\mathbf{r} \label{c41}%
\end{equation}
Eq. (\ref{c41}) shows that we have reduced the initial disordered problem to a
regular model with an $\psi^{4}$ interaction. Of course, it does not help to
solve it exactly but now we can use approximations well developed in theory of
interacting particles.

The Lagrangian $L$ in Eq. (\ref{c41}) is similar to those studied in field
theory. Let us remark that at $\omega=0$ the Lagrangian is invariant under
rotations of the supervectors in the superspace because it depends on the
square of ``length'' only. The frequency $\omega$ violates this symmetry and,
if one uses an analogy with spin models, plays the role of an ``external
field''. The violation of supersymmetry by the frequency is due to the fact
that Eq. (\ref{c41}) is written for the product $G^{R}G^{A}$ (the presence of
the matrix $\Lambda$ is a direct consequence of this). The symmetry would not
be violated in the corresponding integrals for $\left\langle G^{R}%
G^{R}\right\rangle $ and $\left\langle G^{A}G^{A}\right\rangle .$ The
averaging of the simpler Lagrangian corresponding to the density of states
(one averaged Green function) results also in a model with the interaction
$\psi^{4},$ but the supersymmetry in this case is not violated. The diffusion
modes discussed previously exist only as a result of the violation of the
supersymmetry (Goldstone modes).

\subsection{Spontaneous breaking of the symmetry and Goldstone modes.
Non-linear $\sigma$-model.}

It is clearly not possible to calculate any correlation function with the
Lagrangian $L,$ Eq. (\ref{c41}), exactly (except for the case of the
one-dimensional chain or the Bethe lattice, where one can write recurrence
equations. Further calculations will be performed in the limit of large mean
free times $\tau,$ which correspond to a weak interaction in the Lagrangian
$L.$ However, the use of the standard perturbation theory as we have seen is
impossible even in this limit because of the existence of the diffusion modes
and so one should try to use non-perturbative approaches.

One of the standard approaches used for such a type of theories is the mean
field approximation. According to this scheme one simplifies the $\psi^{4}$
interaction replacing pairs $\psi\psi$ by their averages. For the interaction
$\psi^{4}$ there can be six different pairings, which can be written as
follows%
\begin{equation}
L_{int}=\frac{1}{4\pi\nu\tau}\int\left(  \bar{\psi}\psi\right)  ^{2}%
d\mathbf{r}\rightarrow L_{1}+L_{2}+L_{3}, \label{c42}%
\end{equation}%
\[
L_{1}=\frac{1}{4\pi\nu\tau}\sum_{\alpha,\beta}\int2\left\langle \bar{\psi
}_{\alpha}\psi_{\alpha}\right\rangle _{eff}\bar{\psi}_{\beta}\psi_{\beta
}d\mathbf{r,}%
\]%
\[
L_{2}=\frac{1}{4\pi\nu\tau}\sum_{\alpha,\beta}\int2\bar{\psi}_{\alpha
}\left\langle \psi_{\alpha}\bar{\psi}_{\beta}\right\rangle _{eff}\psi_{\beta
}d\mathbf{r,}%
\]%
\[
L_{3}=\frac{1}{4\pi\nu\tau}\sum_{\alpha,\beta}\int\left(  \left\langle
\bar{\psi}_{\alpha}\bar{\psi}_{\beta}\right\rangle _{eff}\psi_{\beta}%
\psi_{\alpha}+\bar{\psi}_{\alpha}\bar{\psi}_{\beta}\left\langle \psi_{\beta
}\psi_{\alpha}\right\rangle \right)  d\mathbf{r}%
\]
In Eqs.(\ref{c42}), the symbol
\[
\left\langle ...\right\rangle _{eff}=\int\left(  ...\right)  \exp\left(
-L_{eff}\right)  D\psi
\]
stands for the functional averaging with the effective Lagrangian
$L_{eff}=L_{0}+L_{1}+L_{2}+L_{3},$ where $L_{0}$ is the quadratic part of the
Lagrangian $L$ in Eq. (\ref{c41}), and $\alpha$ and $\beta$ stand for the
components of the supervectors $\bar{\psi}$ and $\psi.$

In fact the terms in $L_{2}$ and $L_{3}$ are equal to each other. The average
$\left\langle \bar{\psi}_{\alpha}\psi_{\alpha}\right\rangle $ renormalizes the
energy $\varepsilon$ and is not important. The averages in $L_{2}$ and $L_{3}$
can be both commuting and anticommuting variables, depending on the subscripts
$\alpha$ and $\beta.$ The final Lagrangian $L_{eff}$ takes the form%
\begin{equation}
L_{eff}=\int\left[  -i\bar{\psi}\left(  \tilde{H}+\frac{1}{2}\left(
\omega+i\delta\right)  \Lambda+\frac{iQ}{2\tau}\right)  \psi d\mathbf{r}%
\right]  \label{c43}%
\end{equation}
with the $8\times8$ supermatrix $Q$ satisfying the following self-consistency
equation%
\begin{equation}
Q=\frac{2}{\pi\nu}\left\langle \psi\bar{\psi}\right\rangle _{eff} \label{c44}%
\end{equation}
Calculating the Gaussian integral in Eq. (\ref{c44}) with the help of
Eq.(\ref{c29}) we obtain%
\begin{equation}
Q=\frac{1}{\pi\nu}\int g_{0}\left(  \mathbf{p}\right)  \frac{d^{d}\mathbf{p}%
}{\left(  2\pi\right)  ^{d}} \label{c45}%
\end{equation}%
\begin{equation}
g_{0}\left(  \mathbf{p}\right)  =i\left(  \varepsilon\left(  \mathbf{p}%
\right)  +\frac{\omega+i\delta}{2}-\varepsilon+\frac{1}{2}\left(
\omega+i\delta\right)  \Lambda+\frac{iQ}{2\tau}\right)  ^{-1} \label{c46}%
\end{equation}
The integral over the momenta $\mathbf{p}$ in Eq. (\ref{c46}) has both a real
and an imaginary part. As concerns the imaginary part the main contribution
comes from the region $\left|  \varepsilon\left(  \mathbf{p}\right)
\mathbf{-}\varepsilon\right|  \gg\tau^{-1},\omega$ and, therefore, is
proportional to the unit matrix $\mathbf{1.}$ This contribution leads to a
small renormalization of the energy $\varepsilon.$ Assuming that this energy
has already been renormalized we can forget about the imaginary part of $Q$
and concentrate on the real part. The main contribution to the real part of
Eq. (\ref{c46}) comes from the region $\left|  \varepsilon\left(
\mathbf{p}\right)  -\varepsilon\right|  \sim\tau^{-1}\ll\varepsilon.$
Introducing the variable $\xi=\varepsilon\left(  \mathbf{p}\right)
-\varepsilon$ we can rewrite Eq. (\ref{c46}) in the form
\begin{equation}
Q=\frac{i}{\pi}\int_{-\infty}^{\infty}\left(  \xi+\frac{1}{2}\left(
\omega+i\delta\right)  \Lambda+\frac{iQ}{2\tau}\right)  ^{-1}d\xi\label{c47}%
\end{equation}
Eq. (\ref{c47}) determines the contribution to the real part of $Q$ only and
has at $\omega\neq0$ one evident solution%
\begin{equation}
Q=\Lambda,\text{ }\omega\neq0 \label{c48}%
\end{equation}
However, putting $\omega=0$ we see that any $Q$ of the form
\begin{equation}
Q=V\Lambda\bar{V} \label{c49}%
\end{equation}
where $V$ is an arbitrary unitary supermatrix, $V\bar{V}=1,$ satisfies Eq.
(\ref{c47}). Here the symbol of conjugation $``-"$ for an arbitrary matrix $A$
means the following
\begin{equation}
\bar{A}=CA^{T}C^{T} \label{c49a}%
\end{equation}
with $C$ from Eqs. (\ref{c35}, \ref{c37}). The supermatrix $Q$ is self
conjugate, $Q=\bar{Q}.$

The degeneracy of the ground state, Eq. (\ref{c49}), leads to the existence of
the low-lying Goldstone modes, and their contribution to physical quantities
should be taken into account properly. These Goldstone modes are just the
diffusion modes discussed in the Introduction. In the language of spin models
the degeneracy of the solution, Eq. (\ref{c49}) is equivalent to the
degeneracy due to an arbitrary spin direction.

Substituting the mean field solution, Eq. (\ref{c49}), into the effective
Lagrangian into Eq. (\ref{c43}) we obtain zero, which shows that this
approximation is not sufficient. In order to describe a contribution of the
diffusion modes we have to take into account slow variations of the
supermatrix $Q$ in space. This can be done assuming that the supermatrix $V$
is a slow function of the coordinates $\mathbf{r}$ and expanding in the
gradients of $V$ (or, equivalently, $Q$).

As a result of the expansion in the gradients and small frequencies one can
obtain a functional $F\left[  Q\right]  $ describing slow variations of the
supermatrix $Q$ in space%
\begin{equation}
F\left[  Q\right]  =\frac{\pi\nu}{8}STr\int\left[  D_{0}\left(  \nabla
Q\right)  ^{2}+2i\left(  \omega+i\delta\right)  \Lambda Q\right]  d\mathbf{r}
\label{c50}%
\end{equation}
where $D_{0}$ is the classical diffusion coefficient, Eq. (\ref{a8}), and the
supermatrix $Q$ is described by Eq. (\ref{c49}). The free energy functional,
Eq. (\ref{c50}), has the form of a non-linear $\sigma$-model.

In order to calculate, e.g. the level-level correlation function $R\left(
\omega\right)  ,$ Eq. (\ref{c33}), one should express it in terms of a
functional integral over the supermatrix $Q.$ As a result, this function takes
the form%
\begin{equation}
R\left(  \omega\right)  =\frac{1}{2}-\frac{1}{2V^{2}}\operatorname{Re}\int
Q_{11}^{11}\left(  \mathbf{r}\right)  Q_{11}^{22}\left(  \mathbf{r}^{\prime
}\right)  \exp\left(  -F\left[  Q\right]  \right)  DQd\mathbf{r}%
d\mathbf{r}^{\prime} \label{c52}%
\end{equation}
where $V$ is the volume of the system. In Eq. (\ref{c52}) the superscripts of
$Q$ enumerate ``retarded-advanced blocks'' , the subscripts relate to the
matrix elements within these blocks.

Eqs. (\ref{c50}, \ref{c52}) show that the calculation of physical quantities
for disordered systems can be reduced to study of a supermatrix non-linear
$\sigma$-model. There is a variety of physical problems that can be solved by
considering this model in different dimensionalities. To some extent, the
problem of the level statistics in a limited volume is the simplest one
because it corresponds to the zero-dimensional $\sigma$-model (no dependence
of $Q$ on the coordinates).

Adding magnetic or spin orbit interactions results in additional ``external
fields'' partially breaking the symmetry. If these interactions are not very
weak they simply cut some degrees of freedom. As a result, one comes again to
Eq. (\ref{c50}) but with a reduced symmetry of the supermatrix $Q.$

There can be three different types of the symmetries:

1. In the absence of both magnetic and spin-orbital interactions the system is
invariant under time reversal and spatial inversion. It will be called
``orthogonal'' because, as we will see, one comes in small particles to the
Wigner-Dyson statistic for the orthogonal ensemble.

2. In the presence of magnetic interactions the time reversal symmetry is
violated and this will be called unitary ensemble.

3. In the absence of magnetic interactions but in the presence of spin-orbital
ones, one obtains the symplectic ensemble.

These three ensembles lead to quite different results in different situations
and each of them should be considered separately.

\subsection{Level statistics in a limited volume.}

Eq. (\ref{c49}) specifies a general form for the $8\times8$ supermatrix $Q.$
In other words, it obeys the constraint
\begin{equation}
Q^{2}=1 \label{c53}%
\end{equation}

However, this constraint is not sufficient to determine unambiguously the
precise structure of $Q$ because Eq. (\ref{c53}) could correspond both to
rotations on a sphere and on a hyperboloid. It turns out that the symmetry of
the supermatrix $Q$ is more complex than those corresponding to these two
possibilities. The supermatrix $Q$ consists of two parts: one describing
rotations on the sphere and the other on the hyperboloid, such that the group
of rotations of $Q$ is a mixture of compact and non-compact groups of
rotations \cite{ebook}. These two parts are glued by anticommuting elements.
We can describe the explicit form of the supermatrix $Q$ writing it in the
form%
\begin{equation}
Q=UQ_{0}\bar{U}, \label{c54}%
\end{equation}
where
\begin{equation}
U=\left(
\begin{array}
[c]{cc}%
u & 0\\
0 & v
\end{array}
\right)  \label{c55}%
\end{equation}
and $\bar{u}u=1,$ $v\bar{v}=1.$ All Grassmann variables are included in the
$4\times4$ supermatrices $u$ and $v.$ These matrices contain also some phases.
Their form is simple and not very important for our discussion. More details
can be found in Ref. \cite{ebook}.

The form of $Q_{0}$ is more interesting and can be written as%
\begin{equation}
Q_{0}=\left(
\begin{array}
[c]{cc}%
\cos\hat{\theta} & i\sin\hat{\theta}\\
-i\sin\hat{\theta} & -\cos\hat{\theta}%
\end{array}
\right)  ,\text{ }\hat{\theta}=\left(
\begin{array}
[c]{cc}%
\hat{\theta}_{11} & 0\\
0 & \hat{\theta}_{22}%
\end{array}
\right)  \label{c56}%
\end{equation}
The $2\times2$ matrices are different for different classes of the symmetry.

We see that the symmetry of $Q_{0}$ corresponds to a group rotations on both
the sphere and the hyperboloid. The explicit form of the matrices $\hat
{\theta}_{11}$ and $\hat{\theta}_{22}$ can be expressed as follows%
\begin{equation}
\hat{\theta}_{11}=\left(
\begin{array}
[c]{cc}%
\theta & 0\\
0 & \theta
\end{array}
\right)  ,\text{ }\hat{\theta}_{22}=i\left(
\begin{array}
[c]{cc}%
\theta_{1} & \theta_{2}\\
\theta_{2} & \theta_{1}%
\end{array}
\right)  \label{c57}%
\end{equation}
with $0<\theta<\pi,$ $\theta_{1}>0,$ $\theta_{2}>0$ for the orthogonal
ensemble,%
\begin{equation}
\hat{\theta}_{11}=\left(
\begin{array}
[c]{cc}%
\theta & 0\\
0 & \theta
\end{array}
\right)  ,\text{ }\hat{\theta}_{22}=i\left(
\begin{array}
[c]{cc}%
\theta_{1} & 0\\
0 & \theta_{1}%
\end{array}
\right)  \label{c58}%
\end{equation}
with $0<\theta<\pi,$ $\theta_{1}>0$ for the unitary ensemble, and%
\begin{equation}
\hat{\theta}_{11}=\left(
\begin{array}
[c]{cc}%
\theta_{1} & \theta_{2}\\
\theta_{2} & \theta_{1}%
\end{array}
\right)  ,\text{ }\hat{\theta}_{22}=i\left(
\begin{array}
[c]{cc}%
\theta & 0\\
0 & \theta
\end{array}
\right)  \label{c59}%
\end{equation}
for the symplectic one.

Eqs. (\ref{c50},\ref{c52}, \ref{c56}-\ref{c59}) specify the non-linear
supermatrix $\sigma$-model. What remains to do for the level-level correlation
function, Eq. (\ref{c52}), is to calculate the functional integral over $Q.$
Many of the physical quantities can also be expressed in a form of a
correlation function of the supermatrices $Q$ with the free energy functional
$F\left[  Q\right]  ,$ Eq. (\ref{c50}).

Let us consider now the level-level correlation function in a limited volume.
The functional integral, Eqs. (\ref{c50}, \ref{c52}) can be considerably
simplified in the limit of small frequencies. In a finite volume one can
expand the supermatrix $Q$ in Fourier series. Then, it is not difficult to
understand from Eq. (\ref{c50}) that for $\omega\ll D_{0}/L^{2},$ where $L$ is
the sample size, only the zero space harmonics is essential. In this case one
can integrate over the supermatrices $Q$ not varying in space and the integral
for the function $R\left(  \omega\right)  $, Eq. (\ref{c52}), becomes just a
definite integral over several variables. This integral takes the following
form%
\begin{equation}
R\left(  \omega\right)  =\frac{1}{2}-\frac{1}{2}\operatorname{Re}\int
Q_{11}^{11}Q_{11}^{22}\exp\left(  -F_{0}\left[  Q\right]  \right)  dQ
\label{d1}%
\end{equation}
where $F_{0}\left[  Q\right]  $ takes the form%
\begin{equation}
F_{0}\left[  Q\right]  =\frac{i\pi\left(  \omega+i\delta\right)  }{4\Delta
}STr\left(  \Lambda Q\right)  \label{d2}%
\end{equation}
One can say that Eqs. (\ref{d1}, \ref{d2}) determine a zero dimensional
non-linear $\sigma$-model. In general, it is clear that the dimensionality of
the $\sigma$-model is determined at not very high temperatures by the geometry
of the sample. For example, the one dimensional $\sigma$-model describes a
long wire of a finite thickness.

In order to calculate the integral over $Q$ in Eq. (\ref{d1}) it is very
convenient to use Eqs. (\ref{c56}, \ref{c59}). We see immediately that the
supermatrix $U,$ Eq. (\ref{c55}), drops out from $F_{0}\left[  Q\right]  $
entering the pre-exponential in Eq. (\ref{d1}) only. This allows us to
integrate first over the elements of $U,$ and reduce the integral to the
variables $\hat{\theta}_{11},$ $\hat{\theta}_{22}$ only. The integration over
the Grassmann variables is, according to Eq. (\ref{c4}), a very simple task.
Actually, one has to calculate also Jacobians arising when changing from the
integration over $Q$ to the integration over the ``eigenvalues'' $\lambda,$
$\lambda_{1},$ $\lambda_{2}.$ As a result, one comes to the following
integrals for all three ensembles%
\begin{equation}
R_{orth}\left(  \omega\right)  =1+\operatorname{Re}\int_{1}^{\infty}\int
_{1}^{\infty}\int_{-1}^{1}\frac{\left(  \lambda_{1}\lambda_{2}-\lambda\right)
^{2}\left(  1-\lambda^{2}\right)  }{\left(  \lambda_{1}^{2}+\lambda_{2}%
^{2}+\lambda^{2}-2\lambda\lambda_{1}\lambda_{2}-1\right)  ^{2}} \label{d3}%
\end{equation}%
\[
\times\exp\left[  i\left(  x+i\delta\right)  \left(  \lambda_{1}\lambda
_{2}-\lambda\right)  \right]  d\lambda_{1}d\lambda_{2}d\lambda
\]%

\begin{equation}
R_{unit}\left(  \omega\right)  =1+\frac{1}{2}\operatorname{Re}\int_{1}%
^{\infty}\int_{-1}^{1}\exp\left[  i\left(  x+i\delta\right)  \left(
\lambda_{1}-\lambda\right)  \right]  d\lambda d\lambda_{1} \label{d4}%
\end{equation}%
\begin{equation}
R_{sympl}\left(  \omega\right)  =1+\operatorname{Re}\int_{1}^{\infty}\int
_{0}^{1}\int_{-1}^{1}\frac{\left(  \lambda-\lambda_{1}\lambda_{2}\right)
^{2}\left(  \lambda^{2}-1\right)  }{\left(  \lambda^{2}+\lambda_{1}%
^{2}+\lambda_{2}^{2}-2\lambda_{1}\lambda_{2}\lambda-1\right)  ^{2}} \label{d5}%
\end{equation}%
\[
\times\exp\left[  i\left(  x+i\delta\right)  \left(  \lambda-\lambda
_{1}\lambda_{2}\right)  \right]  d\lambda_{1}d\lambda_{2}d\lambda
\]
We see that the integrals over the supermatrix $Q$ can be reduced to integrals
over the ``eigenvalues'' $\lambda,$ $\lambda_{1},$ $\lambda_{2}.$ Depending on
the ensemble one obtains twofold or threefold integrals. Calculation of such
integrals is a much simpler task than calculation of integrals over a large
number $N$ of variables encountered in RMT\cite{mehta}. Although the reduction
to the integrals over the eigenvalues has been carried out for the level-level
correlation function only, the corresponding manipulations for studing other
physical quantities are the same. The only thing that remains to be done when
calculating different physical quantities is to write a proper pre-exponential
and carry out integration over the elements of supermatrices $u$ and $v$
entering the pre-exponential only. Then one obtains integrals over the
variables $\lambda,$ $\lambda_{1},$ $\lambda_{2}$ analogous to those in Eqs.
(\ref{d3}, \ref{d4}, \ref{d5}).

The integration over $\lambda$ and $\lambda_{1}$ in Eq. (\ref{d4}) for the
unitary ensemble is simple. At first glance, the integrals for the orthogonal
and symplectic ensembles look terrifying. However, they can be calculated by
Fourier transforming the integrals from the frequencies $\omega$ to the real
time $t.$ As a result, one comes to the Wigner-Dyson formulae, Eqs.
(\ref{a4}-\ref{a6}), which demonstrates that the level-level correlation
function for a small metal particle is really the same as that given by the
RMT. This is how one proves the relevance of the RMT for disordered systems
\cite{efetov1}. Actually, the model of the disordered metal was the first
microscopic model for which the Wigner-Dyson statistics had been proven.

Of course, this is possible for not very high frequencies $\omega\ll
D_{0}/L^{2}.$ In the opposite limit $\omega\gg D_{0}/L^{2},$ the situation is
no longer zero dimensional but one can do perturbation theory in diffusion
modes. This calculation was done in Ref. \cite{altshk}.

One can also come to the zero dimensional $\sigma$-model starting from the
Gaussian distribution for the random matrices, Eq. (\ref{a1}), in the limit of
large $N$. This was done in the review, Ref. \cite{vwz}. Therefore, the
supersymmetry method can be considered as an alternative to the method of the
orthogonal polynomials\cite{mehta} in RMT.

Using the non-linear $\sigma$-model, Eq. (\ref{c50}), one can consider thick
disordered wires. This corresponds to the one dimensional $\sigma$-model. In
this case one can use a transfer matrix technique that allows to reduce
calculation of a one dimensional functional integral to solving of an
effective ``Schr\"{o}dinger equation''. The exact solution found for this
model\cite{el} proves localization of all states (vanishing of the
conductivity) for any arbitrarily weak disorder. In the language of random
matrices this quasi-one-dimensional model corresponds to a model of random
banded matrices\cite{fm}.

The exact solution can also be found for a model with disordered grains
connected in a such a way that they constitute a Bethe lattice. For this
model, using recursion relations one can write a non-linear integral
equation\cite{eb}. It was demonstrated that within the model on the Bethe
lattice one could have a metal-insulator transition with a very unusual
critical behavior. The model on the Bethe lattice has been shown to be
equivalent to models of certain sparse random matrices \cite{mf}. Sparse
matrices are relevant for description of diluted spin models, some
combinatorial optimization problems\cite{mp} and other interesting systems.

Properties of two dimensional disordered metals can be studied using a
renormalization group scheme \cite{wegner,elk,sw,efetov1,ereview}

\section{Wave functions fluctuations in a finite volume. Multifractality.}

\subsection{General definitions.}

In this Section statistics of wave functions is discussed. Investigation of
wave functions is complimentary to study of energy levels. In the language of
random matrices, energy levels correspond to eigenvalues of the matrices
whereas the wave functions relate to eigenvectors.

Study of wave functions has become popular in condensed matter physics not
long ago with the development of mesoscopic physics. One can study, e.g.,
electron tunneling through so called quantum dots, which are actually small
quantum wells. At low temperatures one can reach a resonance tunnelling regime
when the electron tunnel via one resonance level. In this case the tunnelling
amplitude is very sensitive to the wave function of the resonance state.

Wave functions can also be measured in optical and acoustic resonators where
they are electromagnetic or sound waves, respectively.

We start with the standard Schr\"{o}dinger equation
\begin{equation}
H\phi_{\alpha}\left(  r\right)  =\varepsilon_{\alpha}\phi_{\alpha}\left(
r\right)  \label{e1}%
\end{equation}
that determines the eigenenergies $\varepsilon_{\alpha}$ and eigenfunctions
$\phi_{\alpha}\left(  r\right)  .$

We assume that a finite volume $V$ is considered, such that the spectrum of
the energies $\varepsilon_{\alpha}$ is discrete.

The full statistics of the wave functions $\phi_{\alpha}\left(  r\right)  $ at
a given point $r$ can be described by the following distribution function $f$%
\begin{equation}
f\left(  t\right)  =\Delta\left\langle \sum_{\alpha}\delta\left(  t-\left|
\phi_{\alpha}\left(  r\right)  \right|  ^{2}\right)  \delta\left(
\varepsilon-\varepsilon_{\alpha}\right)  \right\rangle \label{e2}%
\end{equation}
The function $f\left(  t\right)  ,$ Eq. (\ref{e2}), gives the probability that
the square of the absolute value of the wave function (intensity) at the point
$r$ and energy $\varepsilon$ is equal to $t.$ The distribution function
$f\left(  t\right)  $ and the wave functions $\phi_{\alpha}\left(  r\right)  $
are assumed to be properly normalized such that
\begin{equation}
t_{0}=1,\text{ }t_{1}=V^{-1} \label{e3}%
\end{equation}
where $t_{n}\left(  V\right)  $ are coefficients of the so called inverse
participation ratio%
\begin{equation}
t_{n}=\int_{0}^{\infty}t^{n}f\left(  t\right)  dt=\Delta\left\langle
\sum_{\alpha}\left|  \phi_{\alpha}\left(  r\right)  \right|  ^{2n}%
\delta\left(  \varepsilon-\varepsilon_{\alpha}\right)  \right\rangle
\label{e4}%
\end{equation}

These coefficients indicate very sensitively the degree of localization of
states through their dependence $t_{n}\left(  V\right)  $ on the volume of the
system. In a pure metal or a ballistic chaotic box where the wave functions
extend over the whole system one has%
\begin{equation}
t_{n}\varpropto V^{-n} \label{e5}%
\end{equation}
If disorder makes the localization length $L_{c},$ at which the typical wave
functions decay, is much shorter than the sample size $L\sim V^{1/d},$ the
coefficients $t_{n}$ are insensitive to $L.$ However, a very interesting
information about the development of localization can be gained through an
analysis of $t_{n}\left(  V\right)  $ for small samples with $L<L_{c}.$

As soon as the localization length $L_{c}$ exceeds the sample size, any length
scale disappears and, in the language of the coefficients $t_{n},$ this is
described as
\begin{equation}
Vt_{n}\varpropto L^{-\tau\left(  n\right)  },\text{ }\tau\left(  n\right)
=\left(  n-1\right)  d^{\ast}\left(  n\right)  \label{e6}%
\end{equation}
where $d^{\ast}\left(  n\right)  $ may differ from the physical dimension $d$
of the system and be a function of $n.$ This function gives the values of the
\textit{fractal dimensions }$d^{\ast}\left(  n\right)  $\textit{\ }for
each\textit{\ }$n.$\textit{\ }

If the behavior of the wave function is described by Eq. (\ref{e5}), the
fractal dimension $d^{\ast}$ coincides with the physical dimension $d.$ We
will see that the fractal dimension of a system obeying the RMT coincides with
the physical dimension. In such a situation, although the amplitude
fluctuations are possible, they are not very strong.

Once we assume that the envelope of a typical wave function at a length scale
shorter than $L_{c}$ obeys a power law $\phi\left(  r\right)  \varpropto
r^{-\mu}$ with a single fixed exponent $\mu<d/2,$ the set of the coefficients
$t_{n}$ reveals $d^{\ast}=d-2\mu$ different from $d$ but the same for all
$n>d/\left(  2\mu\right)  .$ This is when one speaks of fractal behavior with
the \textit{fractal dimension} $d^{\ast}.$

If $d^{\ast}\left(  n\right)  $ is not a constant, that signals a more
sophisticated structure of the wave functions. They can be imagined as
splashes of multiply interfering waves at different scales and with various
amplitudes, and possibly, self-similarity characterized by a relation between
the amplitude $t$ of the local splash of the wave function and the exponent
$\mu\left(  t\right)  $ of the envelope of its extended power law tail.

We will see below that the multifractality of the wave functions of two
dimensional weakly disordered conductors is the most general property of these
systems as soon as the sample size $L$ does not exceed the localization length
$L_{c}$ \cite{fe}.

\subsection{Porter-Thomas distribution.}

Before starting more complicated calculations I would like to present here
what is known about the distribution of wave functions from nuclear physics
(see, e.g. Refs. \cite{porter,brody}) where it was studied for description of
level width fluctuations in neutron scattering. The wave function fluctuations
are obtained there again from the RMT.

To start the calculation one should choose and arbitrary basis of
eigenfunctions $\rho_{m}\left(  \mathbf{r}\right)  $ and expand the function
$\phi_{n}\left(  \mathbf{r}\right)  $ in this basis%
\begin{equation}
\phi_{n}\left(  \mathbf{r}\right)  =V^{-1/2}\sum_{m=1}^{\infty}a_{nm}\rho
_{m}\left(  \mathbf{r}\right)  \label{e7}%
\end{equation}
where $V$ is the volume. It is convenient to truncate the basis to a finite
$N$-dimensional set and take the limit $N\rightarrow\infty$, as is usually
done in the RMT. The main statistical hypothesis is that all coefficients
$a_{mn}$ are uniformly distributed. The only restriction on the coefficients
$\left\{  a_{mn}\right\}  $ is imposed by normalization of the wave functions,
and the probability density $P\left(  \left\{  a_{mn}\right\}  \right)  $ can
be written as
\begin{equation}
\tilde{P}\left(  \left\{  a_{mn}\right\}  \right)  =\frac{2}{\Omega_{N}}%
\delta\left(  \sum_{m=1}^{N}\left|  a_{mn}\right|  ^{2}-1\right)  \label{e8}%
\end{equation}
where $\Omega_{n}$ is the solid angle in $N$ dimensions. Because of the
truncation of the basis the condition of completeness of the basis $\left\{
\rho_{m}\right\}  $ should be written in the form%
\begin{equation}
\sum_{m=1}^{\infty}\rho_{m}^{2}\left(  \mathbf{r}\right)  \equiv\left|
\vec{\eta}\right|  ^{2}=N \label{e9}%
\end{equation}
where $\vec{\eta}$ is an $N$-dimensional vector with components $\left\{
\rho_{m}\right\}  .$ The distribution function of the intensities $W\left(
v\right)  $ at the point $\mathbf{r}$\textbf{\ }is introduced as
\begin{equation}
W\left(  v\right)  =\frac{2}{\Omega_{N}}\int\delta\left(  v-\left|  \vec
{a}\vec{\eta}\right|  ^{2}\right)  \delta\left(  \left|  \vec{a}\right|
^{2}-1\right)  d\vec{a} \label{e10}%
\end{equation}
where the vector $\vec{a}$ is an $N$-dimensional vector with components
$\left\{  a_{mn}\right\}  .$ The distribution function $W\left(  v\right)  ,$
Eq. (\ref{e10}) is defined in such a way that%
\begin{equation}
\int W\left(  v\right)  dv=1 \label{e11}%
\end{equation}
In the unitary ensemble, one should integrate over complex vectors $\vec{a}.$
Integrating first over the component of the vector $\vec{a}$ parallel to the
vector $\vec{\eta}$ and using Eq. (\ref{e9}) one obtains%
\begin{equation}
W\left(  v\right)  =\frac{2\pi}{N\Omega_{N}}\int\delta\left(  \left|  \vec
{a}_{\perp}\right|  ^{2}-\left(  1-\frac{v}{N}\right)  \right)  d\vec{a}
\label{e12}%
\end{equation}
where $\vec{a}_{\perp}$ is the component perpendicular to $\vec{\eta}.$ The
remaining integration in Eq. (\ref{e12}) can be carried out easily. Taking the
limit $N\rightarrow\infty$ one obtains for the unitary ensemble a simple
formula%
\begin{equation}
W\left(  v\right)  =\exp\left(  -v\right)  \label{e13}%
\end{equation}
Computing the integral in Eq. (\ref{e10}) for real vectors $\vec{a}$ and
$\vec{\eta}$ one can obtain the distribution function for the orthogonal
ensemble%
\begin{equation}
W\left(  v\right)  =\frac{1}{\sqrt{2\pi v}}\exp\left(  -\frac{v}{2}\right)
\label{e14}%
\end{equation}
The functions $W\left(  v\right)  $, Eqs. (\ref{e13},\ref{e14}) satisfy the
normalization conditions, Eq. (\ref{e11}). The distribution functions
$W\left(  v\right)  $ are universal and do not depend on details of models for
disorder. The amplitudes $v$ are related to $t$ from the previous section as
$v=Vt.$ The functions $W$ and $f$ are related to each other accordingly. Eqs.
(\ref{e13}, \ref{e14}) are usually referred to as the Porter-Thomas distribution.

\subsection{Non-linear $\sigma$-model and the statistics of wave functions.}

Now we concentrate on the calculation of the distribution function $f\left(
t\right)  ,$ Eq. (\ref{e2}). At first glance, this task does not seem easy. In
the previous section we were able to reduce the level-level correlation
function to a functional integral over $8\times8$ supermatrices $Q.$ It became
possible because the level-level correlation function $R\left(  \omega\right)
,$ Eq. (\ref{a3}), could be expressed in terms of the product of two Green
functions, Eq. (\ref{c33}). Actually, the size $8\times8$ of the supermatrices
is determined by the fact that only two Green functions are needed.

So, in order to calculate the distribution function $f\left(  t\right)  ,$ Eq.
(\ref{e2}), we have to make two necessary steps:

1. To express Eq. (\ref{e2}) in terms of the Green functions.

2. To express products of the Green functions in terms of an integral over
supervectors $\psi$. If we really want to make explicit calculations the
supervectors $\psi$ should not have too many components.

It turns out that both the steps are possible and the necessary number of the
components of the supervector $\psi$ is just $8$ for the orthogonal ensemble
and $4$ for the unitary one.

The step $1$ is done introducing Green functions $G_{\varepsilon\gamma}^{R,A}$
for a system with smeared levels%
\begin{equation}
G_{\varepsilon,\gamma}^{R,A}\left(  \mathbf{r,r}^{\prime}\right)
=\sum_{\alpha}\frac{\phi_{\alpha}\left(  \mathbf{r}\right)  \phi_{\alpha
}^{\ast}\left(  \mathbf{r}\right)  }{\varepsilon-\varepsilon_{\alpha}\pm
\frac{i\gamma}{2}} \label{e15}%
\end{equation}
Then, Eq. (\ref{e2}) can be written as
\begin{equation}
f\left(  t\right)  =\Delta\left\langle \lim_{\gamma\rightarrow0}\sum_{\alpha
}\delta\left(  t-\frac{i\gamma}{2}G_{\varepsilon\gamma}^{R}\left(
\mathbf{r,r}\right)  \right)  \delta\left(  \varepsilon-\varepsilon_{\alpha
}\right)  \right\rangle \label{e16}%
\end{equation}%

\[
=\frac{\Delta}{2\pi}\lim_{\gamma\rightarrow0}\lim_{\beta\rightarrow
0}\left\langle \int\delta\left(  t-\frac{i\gamma}{2}G_{\varepsilon\gamma}%
^{R}\left(  \mathbf{r,r}\right)  \right)  \left(  G_{\varepsilon\beta}%
^{A}\left(  \mathbf{r}^{\prime},\mathbf{r}^{\prime}\right)  -G_{\varepsilon
\beta}^{R}\left(  \mathbf{r}^{\prime},\mathbf{r}^{\prime}\right)  \right)
\right\rangle d\mathbf{r}^{\prime}%
\]
In Eq. (\ref{e16}) one should first take the limit $\beta\rightarrow0$ and
then $\gamma\rightarrow0.$ Since the distribution function $f\left(  t\right)
$ is represented in terms of a function of only two Green functions at two
points $\mathbf{r}$ and $\mathbf{r\prime}$ one can express it in terms of an
integral over $8$-component supervectors $\psi$ using the Wick theorem. The
corresponding Lagrangian is the same as the one for the level-level
correlation function, Eq. (\ref{c41}), provided one replaces the frequency
$\omega$ by the level width $\gamma.$

The derivation of the corresponding $\sigma$-model is standard and one comes
to the following expression for the distribution function $f\left(  t\right)
$%
\begin{align}
f\left(  t\right)   &  =\lim_{\gamma\rightarrow0}\int\int STr\left(  \pi
_{b}^{\left(  1\right)  }Q\left(  \mathbf{r}\right)  \right)  \delta\left(
t-\frac{\pi\nu\gamma}{4}STr\left(  \pi_{b}^{\left(  2\right)  }Q\left(
\mathbf{r}_{o}\right)  \right)  \right) \label{e17}\\
&  \times\left(  -F\left[  Q\right]  \right)  DQ\frac{d\mathbf{r}}%
{4V}\nonumber
\end{align}
where the free energy functional $F\left[  Q\right]  $ has the form%
\begin{equation}
F=\frac{\pi\nu}{8}\int STr\left[  D_{0}\left(  \nabla Q\right)  ^{2}%
-\gamma\Lambda Q\left(  \mathbf{r}\right)  \right]  \label{e18}%
\end{equation}
and $\mathbf{r}_{o}$ is the ``observation point''. When the system can be
described by the zero dimensional $\sigma$- model the distribution function
$f\left(  t\right)  $ does not depend on $\mathbf{r}_{o}.$ However, beyond the
$0D$ approximation, this function can also be a function of the coordinates.
The matrices $\pi_{b}^{\left(  1,2\right)  }$ in Eq. (\ref{e17}) select from
the supermatrix $Q$ its boson-boson sector and have the form%
\begin{equation}
\pi_{b}^{\left(  1\right)  }=\left(
\begin{array}
[c]{cc}%
\pi_{b} & 0\\
0 & 0
\end{array}
\right)  ,\text{ }\pi_{b}^{\left(  2\right)  }=\left(
\begin{array}
[c]{cc}%
0 & 0\\
0 & \pi_{b}%
\end{array}
\right)  ,\text{ }\pi_{b}=\left(
\begin{array}
[c]{cc}%
0 & 0\\
0 & 1
\end{array}
\right)  \label{e19}%
\end{equation}
As we have discussed previously, the $\sigma$-model is noncompact, Eqs.
(\ref{c56}-\ref{c59}). Therefore, in order to avoid divergent integrals over
the variables $\hat{\theta}_{22},$ we must calculate the integrals keeping
$\gamma$ finite. The limit $\gamma\rightarrow0$ can be taken only at the end
of the calculations. However, it is not very convenient to keep an additional
free parameter, and it is better to get rid of the parameter $\gamma$ at an
earlier stage. This can be done by integrating over the zero space harmonics
of $Q$ in the very beginning of the calculations.

To carry out this procedure one should represent the supermatrix $Q\left(
\mathbf{r}\right)  $ in the form of Eq. (\ref{c49}) and change the variables
of integration $V\left(  \mathbf{r}\right)  $ to $\tilde{V}\left(
\mathbf{r}\right)  $ as $V\left(  \mathbf{r}\right)  =V\left(  \mathbf{r}%
_{o}\right)  \tilde{V}\left(  \mathbf{r}\right)  .$ This leads to
supermatrices $\tilde{Q}:$ $Q\left(  \mathbf{r}\right)  =V\left(
\mathbf{r}_{o}\right)  \tilde{Q}\left(  \mathbf{r}\right)  \bar{V}\left(
\mathbf{r}_{o}\right)  .$ In terms of the new variables $\tilde{V}\left(
\mathbf{r}\right)  $ and $\tilde{Q}\left(  \mathbf{r}\right)  $ the gradient
term in Eq. (\ref{e18}) preserves its form, but now the condition
\begin{equation}
V\left(  \mathbf{r}_{o}\right)  =1,\text{ }\tilde{Q}\left(  \mathbf{r}%
_{o}\right)  =\Lambda\label{e20}%
\end{equation}
has to be fulfilled. Changing the variables of the integration for all points
$\mathbf{r\neq r}_{o}$ from $Q\left(  \mathbf{r}\right)  $ to $\tilde
{Q}\left(  \mathbf{r}\right)  $ one obtains a new free energy functional that
does not contain $V\left(  \mathbf{r}_{o}\right)  $ or $Q\left(
\mathbf{r}_{o}\right)  .$ These variables enter only the pre-exponential and
the term with $\gamma$, and, hence, the integral over $V\left(  \mathbf{r}%
_{o}\right)  $ can be computed without making approximations. The result of
the integration contains only the variables $\tilde{Q}\left(  \mathbf{r}%
\right)  $ with the boundary condition, Eq. (\ref{e20}). This means that the
reduced $\sigma$-model obtained in this way operates only with relative
variations of the field $Q$ with respect to its value at the observation point.

The limit $\gamma\rightarrow0$ simplifies the computation because the main
contribution to the integral over the variable $\theta_{1o}$ entering the
parametrization, Eqs. (\ref{c56}-\ref{c59}), is from $\cosh\theta_{1o}%
\sim1/\gamma$ (for simplicity we are considering the unitary ensemble but the
final results are similar for all ensembles). After standard manipulations one
can express the distribution function $f\left(  t\right)  $ in the form%
\begin{equation}
f\left(  t\right)  =\frac{1}{V}\frac{d^{2}\Phi\left(  t\right)  }{dt^{2}%
},\text{ }\Phi\left(  t\right)  =\int_{\tilde{Q}\left(  \mathbf{r}_{o}\right)
=\Lambda}\exp\left(  -\tilde{F}\left[  \tilde{Q},t\right]  \right)  D\tilde
{Q}\left(  \mathbf{r}\right)  \label{e21}%
\end{equation}
where the free energy $\tilde{F}\left[  Q,t\right]  $ has the following form
\begin{equation}
\tilde{F}\left[  Q,t\right]  =\frac{1}{8}\int STr\left[  \pi\nu D_{0}\left(
\nabla\tilde{Q}\right)  ^{2}-2t\Lambda\Pi\tilde{Q}\right]  d\mathbf{r}
\label{e22}%
\end{equation}
The matrix $\Pi$ selects from $\tilde{Q}$ its noncompact ``boson-boson'' sector.

If $t$ is not very large one can take into account the zero space harmonics of
$Q$ only. Taking into account Eq. (\ref{e20}) we can just put everywhere
$Q=\Lambda,$ which leads us immediately to the Porter-Thomas distribution, Eq.
(\ref{e13}) (and Eq. (\ref{e14}) for the orthogonal ensemble).

Nonetheless, this would be only an approximate procedure because the value
$\tilde{Q}\left(  \mathbf{r}\right)  =\Lambda$ does not correspond to the
minimum of the functional $\tilde{F}$ when $t\neq0.$ An equation for the
minimum can be found by taking into account the noncompact variable
$\theta_{1}$ under the condition at the boundary and at the observation point%
\begin{equation}
\mathbf{n\nabla}\theta_{1}=0,\text{ }\theta_{1}\left(  \mathbf{r}_{o}\right)
=0 \label{e23}%
\end{equation}
where $\mathbf{n}$ is the unit vector at the boundary and perpendicular to it.

Using Eq. (\ref{e23}) one writes the equation for the extremum solution
$\theta_{t}$ in the form%
\begin{equation}
\Delta_{\mathbf{r}}\theta_{t}=-\frac{t}{\pi\nu D_{0}}\exp\left(  -\theta
_{t}\left(  \mathbf{r}\right)  \right)  \label{e24}%
\end{equation}
where $\Delta_{\mathbf{r}}$ is the Laplacian. The solution $\theta_{t}\left(
\mathbf{r}\right)  $ of Eq. (\ref{e24}) has to be substituted into the energy
functional $\tilde{F},$ Eq. (\ref{e22}), which takes the form%
\begin{equation}
F_{t}=\frac{1}{2}\int\left[  \pi\nu D_{0}\left(  \nabla\theta_{t}\right)
^{2}+2t\exp\left(  -\theta_{t}\right)  \right]  d\mathbf{r} \label{e25}%
\end{equation}
It is remarkable that Eq. (\ref{e24}) for the non-trivial vacuum of the
reduced $\sigma$-model in two dimensions is exactly the Liouville equation
known in the conformal theory of $2D$ quantum gravity \cite{polyakov,seiberg}.
Within this model one has to calculate the functional integral over all
$\theta$ with the free energy functional determined by Eq. (\ref{e25}).
Although this can lead to helpful analogies \cite{tsvelik}, results that might
be anticipated in this way can be used only as intermediate asymptotics.

Eq. (\ref{e24}) for the minimum looks very similar to a saddle point equation
derived in Ref. \cite{mkh} when considering the problem of long-living current
relaxation. However, the non-linear term is different, which leads to
different solutions.

The most interesting is the solution of Eq. (\ref{e24}) in two dimensions
where it can be found exactly. However, the exact solution is somewhat
cumbersome and I write here its asymptotics at distances $r$ much smaller than
the sample size $L$ (but exceeding the mean free path $l$)
\begin{equation}
\exp\left(  -\theta_{t}\right)  \approx\left(  l/r\right)  ^{2\mu} \label{e26}%
\end{equation}
where $\mu$ is a parameter depending on disorder.

With the same accuracy, the free energy of the vacuum state can be
approximated by%
\begin{equation}
F_{t}\approx4\pi^{2}\nu D_{0}\left\{  \mu+\mu^{2}\ln\left(  L/l\right)
\right\}  \label{e27}%
\end{equation}
The parameter $\mu$ can be determined from the following equation%
\begin{equation}
\mu\approx\frac{z\left(  T\right)  }{2\ln\left(  L/l\right)  },\text{ where
}ze^{z}=T\equiv\frac{tV\ln\left(  L/l\right)  }{2\pi^{2}\nu D_{0}} \label{e28}%
\end{equation}

In principle, Eqs. (\ref{e26}-\ref{e28}) determine the distribution function
$f\left(  t\right)  $ for arbitrary $t$ ($\mu$ must remain small, though).
However, analytical expressions can be written only in the limiting cases
$T\ll1$ and $T\gg1.$ In these limits, the distribution function $f\left(
t\right)  ,$ Eq. (\ref{e2}), can be written as\cite{fe}
\begin{equation}
f\left(  t\right)  =AV\left\{
\begin{array}
[c]{cc}%
\exp\left(  -Vt\left[  1-\frac{T}{2}+...\right]  \right)  , & T\ll1\\
\exp\left(  -\frac{\pi^{2}\nu D_{0}}{\ln\left(  L/l\right)  }\ln^{2}T\right)
, & T\gg1
\end{array}
\right.  \label{e29}%
\end{equation}
where $A$ is a normalization constant.

We see from Eq. (\ref{e29}) that at small values of the amplitudes, such that
$T\ll1,$ the distribution function $f\left(  t\right)  $ agrees with the
Porter-Thomas distribution, Eq. (\ref{e13}), thus proving the latter for
disordered systems. In this limit one can make expansion in $T$\cite{fmir}.

At large $t$ ($T\gg1$) the function $f\left(  t\right)  $ has log-normal
asymptotics that is strikingly similar to the asymptotics of the distribution
function of the local density of states or conductances discovered by
Altshuler, Kravtsov and Lerner \cite{akl} who came to this result considering
renormalization of terms high gradients in the $\sigma$-model. Even the
numerical coefficients in the exponentials are the same, although, of course,
the logarithms contain different variables. It appears that the log-normal
form is really universal. The slower decay of the distribution function
$f\left(  t\right)  $ at large $t$ is due to localization effect.
Unfortunately, until now it is not clear how the growth of the high gradient
terms is related to the existence of the non-trivial vacuum considered in this Section.

As concerns the coefficients of the inverse participation ratio $t_{n}$, Eq.
(\ref{e4}), they show the multifractal behavior, Eq. (\ref{e6}). Using Eqs.
(\ref{e24},\ref{e28}) one comes to the following expression for the fractal
dimension $d^{\ast}\left(  n\right)  $%
\begin{equation}
d^{\ast}\left(  n\right)  =2-n\left(  4\pi^{2}\nu D_{0}\right)  ^{-1}
\label{e30}%
\end{equation}
We see that even for a weak disorder the fractal dimension $d^{\ast}\left(
n\right)  $ strongly deviates from $2.$ Of course, it cannot become negative
and this determines the region of the applicability of Eq. (\ref{e30}).

\section{Recent and possible future developments.}

In the preceding sections it was demonstrated how one can come to the random
matrix theory starting from a model of a disordered metallic particle. This
became possible using the supersymmetry method. Actually, to the best of my
knowledge, the model of a disordered metal was the first microscopic model for
which the Wigner-Dyson statistics was confirmed.

Starting from the first works \cite{ge,efetov1} where the relevance of the
Wigner-Dyson theory to the disordered systems was suggested and proven, a huge
number of problems have been attacked using these ideas. Calculations were
performed either assuming that RMT was applicable for the description of small
particles (they are often called ``quantum dots'') and using methods of the
RMT \cite{mehta} or making direct computation starting from a disordered metal
and applying the non-linear supermatrix $\sigma$-model. Reviewing all these
application within several lectures is impossible even though several related
topics are considered in this volume by Boris Altshuler and Jac Verbaarschot.
At this point I can only refer again to recent reviews
\cite{ebook,az,suptrace,fyodorov,gmw,mirlin,ast,asz,been} and apologize in
case if some references are missing here.

The selection of topics of the present lectures was motivated mainly by the
desire to give a feeling of how to calculate within the supersymmetry method
both level and wave function correlations. We have seen that one could obtain
results that agreed in a certain region of parameters with the predictions of
the RMT and, at the same time, go beyond the RMT.

Essential conditions for the derivation of the $\sigma$-model were the absence
of the electron-electron interaction and a sufficiently high concentration of
impurities. For the problem of the level statistics, the latter condition
corresponds to the case when the mean free path $l$ is much smaller than the
sample size. In other words, an electron can scatter many times on the
impurities in the bulk before it reaches a boundary of the sample.

At the same time, the RMT was initially suggested for description of complex
nuclei, where disorder is absent but interactions are strong. A natural
question that can be asked is: Can one prove the relevance of the RMT for
clean or/and interacting systems? Clearly, the supermatrix $\sigma$-model
discussed in the previous sections is not applicable in these situations and
one needs a generalization of this method.

It seems that really new ideas are necessary in order to achieve this goal.
Nevertheless, first steps towards constructing more general schemes have been
done and I want to present here the main ideas of the new approaches.

\subsection{Supersymmetry with interaction.}

From the beginning of the use of the supersymmetry method it was clear that
the method could be applicable for non-interacting particles only. The method
is based on the result of the Gaussian integration, Eq. (\ref{c6}), that gives
$DetA$ instead of the usual $\left(  DetA\right)  ^{-1}.$ Introducing an
interaction results in non-quadratic terms in the Lagrangian. Therefore the
trick with writing Green functions in terms of a Gaussian integral without a
weight denominator does not work anymore. This is the reason why, in contrast
to the replica approach where a proper $\sigma$-model has been derived long
ago \cite{finkelstein,fin1}, introducing an interaction into the supersymmetry
scheme was believed to be impossible.

To some extent, it is true and it is not clear how to include the interaction
into the supersymmetry exactly. However, a weak interaction can really be
included without considerable difficulties \cite{schwiete}. The initial
electron model with the interaction is not supersymmetric and cannot be made
supersymmetric by a transformation. This is why one cannot get rid off the
weight denominator.

The main idea of Ref. \cite{schwiete} is to replace approximately the initial
electron model by an effective supersymmetric model. This is possible for any
disorder in the limit of a weak interaction. The effective model takes into
account the most important (Hartree-Fock type) diagrams for any fixed
configuration of impurities. Then, the derivation of the proper $\sigma$-model
is quite standard.

The resulting non-linear $\sigma$-model resembles very much the replica model
of Finkelstein\cite{finkelstein,fin1} but contains supermatrices and does not
have replica indices. The supermatrices $Q$ contain, in addition to those for
the non-interacting systems, indices for Matsubara frequencies. One should
also write properly spin indices. As a result, the supermatrix non-linear
$\sigma$-model for electron systems with interaction takes the form (unitary
ensemble)%
\begin{align}
F  &  =\frac{\pi\nu}{4}\int d\mathbf{r}\;\mathrm{Str}\big[D(\nabla
Q)^{2}-4EQ]\nonumber\\
&  +\frac{\pi\nu}{4}\int d\mathbf{r}\;\big[\Gamma_{2}Q\gamma_{2}Q-\Gamma
_{1}Q\gamma_{1}Q\big] \label{eq:ssigmamodel1}%
\end{align}
where $\gamma_{1}$ and $\gamma_{2}$ are certain operators acting on the
supermatrices $Q,$ and $\Gamma_{1}$ and $\Gamma_{2}$ are scattering amplitudes
characterizing the interaction (they are different from those obtained for the
replica $\sigma$-model \cite{finkelstein,fin1}. As usual, one has the
constraint $Q^{2}=1$ but now the product of two supermatrices includes
summation also summation over the Matsubara frequencies.

Using the sigma-model Eq. (\ref{eq:ssigmamodel1}) renormalization group
equations of Refs. (\cite{finkelstein,fin1}) have been reproduced in the first
order in the interaction constants and there is a hope to use this model for
non-perturbative calculations.

\subsection{Method of quasiclassical Green functions.}

Although the interaction is included in Eq. (\ref{eq:ssigmamodel1}), it is
written for systems with considerably high concentration of short range
impurities. Studying problems for clean systems or systems with a long range
disorder one needs a different scheme. Statistical properties of clean chaotic
systems are covered in this school by Boris Altshuler and I do not review them
here. Instead, I want to concentrate on calculational schemes.

The saddle point equation (\ref{c47}) is not a good approximation for clean
systems and systems with long range disorder and we cannot follow the same way
as the one used for diffusive models. The method that I want to present now is
based on using quasiclassical Green functions.

Introducing an $8\times8$ matrix function $G\left(  \mathbf{r,r}^{\prime
}\right)  $ as%

\begin{equation}
G(\mathbf{r},\mathbf{r}{^{\prime}})=2\langle\psi(\mathbf{r}){\bar{\psi}%
}(\mathbf{r}{^{\prime}})\rangle_{\psi} \label{f1}%
\end{equation}
we can write in a standard way the following equation for this function
\begin{equation}
\left[  H_{0\mathbf{r}}+U\left(  \mathbf{r}\right)  +\Lambda\frac
{\omega+i\delta}{2}+iJ\left(  \mathbf{r}\right)  \right]  G\left(
\mathbf{r,r}^{\prime}\right)  =i\delta\left(  \mathbf{r-r}^{\prime}\right)
\label{eqgreenfunc}%
\end{equation}
where the subscript $\mathbf{r}$ of $H_{0\mathbf{r}}$ means that the operator
acts on $\mathbf{r.}$ The notations are the same as in Eqs. (\ref{c39},
\ref{c40}), and $J\left(  \mathbf{r}\right)  $ is a source term that allows to
extract more complicated correlation functions.

Conjugating Eq.(\ref{eqgreenfunc})we obtain another equation for the matrix
$G(\mathbf{r},\mathbf{r}^{\prime})$ with the operator $H_{0\mathbf{r}^{\prime
}}$ acting on its second variable%

\begin{equation}
G(\mathbf{r};\mathbf{r}{^{\prime}})\left[  H_{0\mathbf{r}^{\prime}}+U\left(
\mathbf{r}^{\prime}\right)  +\Lambda\frac{\omega+i\delta}{2}+iJ(\mathbf{r}%
{^{\prime}})\right]  =i\delta(\mathbf{r}-\mathbf{r}{^{\prime}})
\label{conjeqgreenfunc}%
\end{equation}

Until now no approximations have been done and Eqs. (\ref{eqgreenfunc},
\ref{conjeqgreenfunc}) are exact. Now we can use the assumption that the
potential $U\left(  \mathbf{r}\right)  $ changes slowly on the wavelength
$\lambda_{F}$. If the mean free path $l$ for the scattering on the random
potential exceeds $\lambda_{F}$ the Green function varies as a function of
$\mathbf{r-r}^{\prime}$ at distances of the order of $\lambda_{F}$ but, at the
same time, is a slow function of $\left(  \mathbf{r+r}^{\prime}\right)  /2$.
The Fourier transform $G_{\mathbf{p}}\left(  \left(  \mathbf{r+r}^{\prime
}\right)  /2\right)  $ of $G$ $\left(  \mathbf{r,r}^{\prime}\right)  $
respective to $\mathbf{r-r}^{\prime}$ has a sharp maximum near the Fermi
surface. In order to cancel large terms we subtract Eq. (\ref{conjeqgreenfunc}%
) from Eq. (\ref{eqgreenfunc}). Using the assumption that the potential
$U\left(  \mathbf{r}\right)  $ is smooth and expanding it in gradients we
obtain in the lowest order%

\begin{equation}
\left[  -\frac{i\mathbf{p\nabla}_{\mathbf{R}}}{m}+i\mathbf{\nabla}%
_{\mathbf{R}}U(\mathbf{R})\frac{\partial}{\partial\mathbf{p}}\right]
G_{\mathbf{p}}(\mathbf{R})+\frac{\omega+i\delta}{2}[\Lambda,G_{\mathbf{p}%
}(\mathbf{R})]+i[J(\mathbf{R}),G_{\mathbf{p}}(\mathbf{R})]=0 \label{f2}%
\end{equation}
where \ $\mathbf{R}=\left(  \mathbf{r}+\mathbf{r}{^{\prime}}\right)  /2$ and
$\left[  ,\right]  $ stands for the commutator. When deriving Eq. (\ref{f2}),
not only the potential $U\left(  \mathbf{r}\right)  $ but also the function
$J\left(  \mathbf{r}\right)  $ was assumed to be smooth.

The dependence of the Green function $G_{\mathbf{p}}\left(  \mathbf{R}\right)
$ on $\mathbf{|p|}$ is more sharp than on other variables. In order to avoid
this sharp dependence we integrate Eq. (\ref{f2}) over $|\mathbf{p|}$. Of
course, this procedure makes a sense for very large samples when the level
discreteness can be neglected.

The most interesting contribution in the integral over $|\mathbf{p|}$ comes
from the vicinity of the Fermi-surface. A contribution given by momenta
considerably different from $p_{F}$ is proportional to the unity matrix an
drops out from Eq. (\ref{f2}).

Introducing the function $g_{\mathbf{n}}\left(  \mathbf{r}\right)  $
\begin{equation}
g_{\mathbf{n}}\left(  \mathbf{r}\right)  =\frac{1}{\pi}\int G_{p\mathbf{n}%
}\left(  \mathbf{r}\right)  d\xi\text{, \ }\xi=\frac{p^{2}-p_{F}^{2}}{2m}
\label{klassaver}%
\end{equation}
where $\mathbf{n}$ is a unite vector pointing a direction on the Fermi
surface, we obtain the final quasiclassical equation

{\large
\begin{equation}
\left(  v_{F}\mathbf{n\nabla}-p_{F}^{-1}\mathbf{\nabla}_{\mathbf{r}}U\left(
\mathbf{r}\right)  \mathbf{\partial}_{\mathbf{n}}\right)  g_{\mathbf{n}%
}\left(  \mathbf{r}\right)  +\frac{i\left(  \omega+i\delta\right)  }%
{2}[\Lambda,g_{\mathbf{n}}\left(  \mathbf{r}\right)  ]-[J,g_{\mathbf{n}}]=0
\label{f2a}%
\end{equation}
} where
\[
\mathbf{\partial}_{\mathbf{n}}=\mathbf{\nabla}_{\mathbf{n}}-\mathbf{n,}\text{
}{\nabla}_{\mathbf{n}}=-[\mathbf{n\times\lbrack n\times}\frac{\partial
}{\partial\mathbf{n}}]]
\]

Eq. (\ref{f2a}) should be complemented by a boundary condition at the surface
of the sample. Considering a closed sample we assume that the current across
the border is equal to zero. This leads the boundary condition at the surface
\begin{equation}
\left.  g_{\mathbf{n}_{\perp}}\left(  \mathbf{r}\right)  \right|
_{surface}=\left.  g_{-\mathbf{n}_{\perp}}\left(  \mathbf{r}\right)  \right|
_{surface} \label{f3}%
\end{equation}
where $\mathbf{n}_{\perp}$ is the component of the vector $\mathbf{n}$
perpendicular to the surface.

Eq. (\ref{f2a}) is similar to an Eilenberger equation written long ago in
superconductivity theory \cite{eilen}. As in the theory of superconductivity,
the solution for the Eq.(\ref{f2a}) satisfies the condition $g_{\mathbf{n}%
}^{2}(\mathbf{r})=\hat{1}.$ Eq. (\ref{f2a}) is written for a non-averaged
potential $U\left(  \mathbf{r}\right)  $ and it is valid also in the absence
of the long range potential.

The quasiclassical equation, Eq. (\ref{f2a}), has been written first by
Muzykantskii and Khmelnitskii, Ref. \cite{mk} who guessed a functional for
which Eq. (\ref{f2a}) is just a condition for an extremum. Then, they
proceeded to work with this functional without estimating fluctuations near
this minimum.

It came later as a surprise that, actually, one could write the proper
solution of Eq. (\ref{f2a}) in terms of a functional integral over
supermatrices exactly\cite{kogan}. The \textit{exact} solution for Eq.
(\ref{f2a}) can be written as%

\begin{align}
&  g_{\mathbf{n}}(\mathbf{r})=Z^{-1}[J]\int_{Q_{\mathbf{n}}^{2}=1}%
Q_{\mathbf{n}}(\mathbf{r})\exp\left(  -\frac{\pi\nu}{2}\Phi_{J}[Q_{\mathbf{n}%
}(\mathbf{r})]\right)  DQ_{\mathbf{n}},\nonumber\\
&  \Phi\lbrack Q_{\mathbf{n}}(\mathbf{r})]=Str\int d\mathbf{r}d\mathbf{n[}%
\Lambda{\bar{T}}_{\mathbf{n}}(\mathbf{r})(v_{F}\mathbf{n\nabla}_{\mathbf{r}%
}\nonumber\\
&  -p_{F}^{-1}\mathbf{\nabla}_{\mathbf{r}}U(\mathbf{r})\mathbf{\nabla
}_{\mathbf{n}})T_{\mathbf{n}}(\mathbf{r})+\left(  \frac{i\left(
\omega+i\delta\right)  }{2}\Lambda-J(\mathbf{r})\right)  Q_{\mathbf{n}%
}(\mathbf{r})],\label{sigmaaverage}\\
&  Q_{\mathbf{n}}(\mathbf{r})=T_{\mathbf{n}}(\mathbf{r})\Lambda{\bar{T}%
}_{\mathbf{n}}(\mathbf{r}),\text{ \ }{\bar{T}}_{\mathbf{n}}(\mathbf{r}%
)T_{\mathbf{n}}\left(  \mathbf{r}\right)  =1\nonumber
\end{align}
In Eq. (\ref{sigmaaverage}), the partition function $Z[J\left(  \mathbf{r}%
\right)  ]$ is
\begin{equation}
Z[J]=\int_{Q_{\mathbf{n}}^{2}=1}\exp\left(  -\frac{\pi\nu}{2}\Phi\lbrack
Q_{\mathbf{n}}(\mathbf{r})]\right)  DQ_{\mathbf{n}} \label{integ}%
\end{equation}
and the integration is performed over the self-conjugate supermatrices
$Q_{\mathbf{n}}=\bar{Q}_{\mathbf{n}}\left(  \mathbf{r}\right)  $ satisfying
the following relation
\begin{equation}
Q_{\mathbf{n}}^{2}\left(  \mathbf{r}\right)  =1 \label{f4}%
\end{equation}
We see that the quasiclassical Green function $g_{\mathbf{n}}\left(  r\right)
$ can be written in the form of a functional integral over supermatrices
$Q_{\mathbf{n}}\left(  \mathbf{r}\right)  $ depending both on the coordinates
$\mathbf{r}$ and the direction of the momentum $\mathbf{n}$ and satisfying the
constraint, Eq. (\ref{f4}). The first term in the free energy functional is
written in terms of the supermatrices $T_{\mathbf{n}}$ rather than
$Q_{\mathbf{n.}}.$ However, it can be written in a form of a
Wess-Zumino-Novikov-Witten term containing the supermatrices $Q$
only\cite{mk}. Writing this term one should introduce an additional coordinate
varying at the interval $\left[  0,1\right]  .$

The model described by the functional $\Phi$, Eqs. (\ref{sigmaaverage}), is
usually referred to as a ``ballistic $\sigma$-model''. The partition function
$Z\left[  J\right]  ,$ Eq. (\ref{integ}), is unity at $J=0$ due to the
supersymmetry and this allows us to average (if necessary) over the smooth
potential $U\left(  \mathbf{r}\right)  $.

The method of quasiclassical Green functions suggested here for a static
external potential does not seem to be restricted by the non-interacting case.
There are indications that it can be generalized to describe clean interacting
systems. Of course, in this case one should write the quasiclassical equations
in time representation because the interaction mixes states with different
energies. Study of interacting systems with this method may be a very
interesting direction of research.

As concerns attempts to prove the Wigner-Dyson statistics for clean
non-interacting systems one can try to start with the functional $\Phi,$ Eq.
(\ref{sigmaaverage}). At first glance, we should simply restrict ourselves
with the integration over $Q$ depending neither on the coordinates, nor on the
momenta. Then, the functional $\Phi$ would contain only the last term and we
would have the zero dimensional $\sigma$-model, which leads immediately to the
WD statistics.

However, a very important question is whether one averages over the potential
$U\left(  \mathbf{r}\right)  $ or puts $U\left(  \mathbf{r}\right)  =0$ and
averages over the spectrum. In the former case one gets after averaging over
$U\left(  \mathbf{r}\right)  $ an additional term in functional $\Phi$
quadratic in gradients. This term leads eventually to a suppression of
non-zero harmonics and one can really obtain the zero-dimensional $\sigma
$-model (see \cite{kogan} and references therein).

The situation with $U\left(  \mathbf{r}\right)  =0$ and averaging over the
energy is more interesting. Everything depends on whether the system is
classically integrable or chaotic. It is just the situation for which the
authors of Ref. \cite{bgs} made their hypothesis.

It turns out that within the model with the functional $\Phi$, Eq.
(\ref{sigmaaverage}), and $U\left(  \mathbf{r}\right)  =0$ one cannot come to
the zero-dimensional $\sigma$-model. There is a common consensus that a
``regularizer'' (see e.g. \cite{aos}) containing something like square of
gradients in coordinates or momenta is necessary in the correct ballistic
$\sigma$-model. Aleiner and Larkin \cite{al} argued that in order to come to
the zero dimensional $\sigma$-model one had to take into account diffraction,
which is clearly absent in the ballistic $\sigma$-model, Eq.
(\ref{sigmaaverage}). They did not manage to include the diffraction in their
calculational scheme microscopically and mimiced it by introducing artificial
quantum impurities that would correspond to the potential $U\left(
\mathbf{r}\right)  $ in Eq. (\ref{sigmaaverage}). This allowed them to come to
the zero dimensional $\sigma$-model, confirm the WD statistics and calculate
corrections to it. It is worth emphasizing that the effective potential
$U\left(  \mathbf{r}\right)  $ was very weak such that the computation was
done in the ballistic regime.

As concerns the real physical diffraction, it cannot be directly included in
the $\sigma$-model using the quasiclassical scheme and a more sophisticated
approach is necessary.

\subsection{Beyond the quasiclassics}

We have seen that the solution of the equations for the Green functions, Eqs.
(\ref{eqgreenfunc}, \ref{conjeqgreenfunc}), can be written in the
quasiclassical approximation in terms of the functional integral over
$8\times8$ supermatrices $Q_{\mathbf{n}}\left(  \mathbf{r}\right)  .$ For
certain problems this approximation is not sufficient and the natural question
is: can we do better than that and find a solution for the Green functions in
terms of a functional integral over supermatrices valid at all distances
including those of the order of the electron wavelength $\lambda_{F}$ ?

This attempt has been undertaken recently in Ref. \cite{superbos}, where an
integral of such a type was suggested for a solution of Eq. (\ref{eqgreenfunc}%
). The idea is rather close to the one known in field theory where it is
called bosonization\cite{boson}. The final expressions obtained in Ref.
(\cite{boson}) are rather complicated and this method, to the best of my
knowledge, has not evolved into an efficient calculational tool.

However, the supersymmetric form of the Green functions considered here seems
to promise more and the derivation is rather simple. I follow here a simpler
derivation of Ref. \cite{kogan1}.

What I want to show now is that the Green function, Eq. (\ref{eqgreenfunc}),
can be represented exactly as an integral over supermatrices $Q\left(
\mathbf{r,r}^{\prime}\right)  $ depending on two coordinates $\mathbf{r}$ and
$\mathbf{r\prime}$
\begin{equation}
G(\mathbf{r},\mathbf{r^{\prime}})=Z^{-1}[J]\int Q(\mathbf{r},\mathbf{r^{\prime
}})\exp\bigl(-\Phi\lbrack Q]\bigr)DQ \label{f5}%
\end{equation}
where $Z[J]$ is a new partition function
\begin{equation}
Z[J]=\int\exp\bigl(-\Phi\lbrack Q]\bigr)DQ \label{f6}%
\end{equation}
and the functional $\Phi\lbrack Q]$ has the form
\begin{align}
\Phi\lbrack Q]=\frac{i}{2}\mathrm{Str}\int &  \left(  H_{0\mathbf{r}}+U\left(
\mathbf{r}\right)  +\frac{\omega+i\delta}{2}\Lambda\right) \nonumber\\
&  \times\delta(\mathbf{r}-\mathbf{r^{\prime}})Q(\mathbf{r},\mathbf{r^{\prime
}})d\mathbf{r}d\mathbf{r^{\prime}} \label{f7}%
\end{align}%
\[
+\frac{1}{2}\mathrm{Str}\ln Q-\frac{1}{2}\mathrm{Str}\int J(\mathbf{r}%
,\mathbf{r^{\prime}})Q(\mathbf{r^{\prime}},\mathbf{r})d\mathbf{r}%
d\mathbf{r^{\prime}}%
\]
where $J\left(  \mathbf{r,r}^{\prime}\right)  $ is a source term. The
structure of the supermatrix $Q(\mathbf{r},\mathbf{r^{\prime}})$ in the
integral Eq.(\ref{f5}) should be the same as that of the Green function
$G(\mathbf{r},\mathbf{r^{\prime}})$, i.e. be the same as of the product
$\psi(\mathbf{r})\bar{\psi}(\mathbf{r^{\prime}})$. In particular, this means
that $Q(\mathbf{r},\mathbf{r^{\prime}})$ is self-conjugated
\begin{equation}
\bar{Q}(\mathbf{r},\mathbf{r^{\prime}})\equiv C\;Q^{T}(\mathbf{r^{\prime}%
},\mathbf{r})C^{T}=Q(\mathbf{r},\mathbf{r^{\prime}}) \label{f8}%
\end{equation}
In order to prove Eq.(\ref{f5}) we write the following identity
\begin{align}
&  -2iZ^{-1}[J]\int\biggl[\int\frac{\delta\exp\left(  -\frac{1}{2}%
\mathrm{Str\ln Q}\right)  }{\delta Q(\mathbf{r^{\prime\prime}},\mathbf{r}%
)}Q(\mathbf{r^{\prime\prime}},\mathbf{r^{\prime}})d\mathbf{r^{\prime\prime}%
}\biggr]\times\nonumber\\
&  \exp\left(  -\frac{i}{2}\mathrm{Str}\left[  \hat{H}_{0\mathbf{r}}%
+\frac{\omega+i\delta}{2}\Lambda+iJ\right]  Q\right)  DQ=\nonumber
\end{align}%
\begin{equation}
=i\delta(\mathbf{r}-\mathbf{r^{\prime}}), \label{f9}%
\end{equation}
and integrate over $Q$ by parts. The derivative $\delta/\delta Q$ should act
now on both $Q$ and the exponential. At this point, the supersymmetry plays a
crucial role. Differentiating first the supermatrix $Q$ we obtain the
supermatrix product $\left(  \delta/\delta Q\right)  Q.$ As the number of the
anticommuting variables in the sum over the matrix elements is equal to the
number of the boson ones and the derivatives have the opposite signs, this
matrix product vanishes. Differentiating the exponential only we come to the
following equation
\begin{align}
&  Z^{-1}[J]\int d\mathbf{r^{\prime\prime}}\left(  \hat{H}_{0\mathbf{r}%
}+U\left(  \mathbf{r}\right)  \frac{\omega+i\delta}{2}\Lambda+iJ\right)
(\mathbf{r},\mathbf{r^{\prime\prime}})\times\nonumber\\
&  \int Q(\mathbf{r^{\prime\prime}},\mathbf{r^{\prime}})\exp\bigl(-\Phi\lbrack
Q]\bigr)DQ=i\delta(\mathbf{r}-\mathbf{r^{\prime}}) \label{f10}%
\end{align}
Eq. (\ref{f10}) proves immediately that the integral Eq.(\ref{f5}) does
satisfy Eq.(\ref{eqgreenfunc}) and we have really the alternative
representation of the Green function in terms of an integral over the
supermatrices $Q$.

Making the Fourier transform $Q\left(  \mathbf{r}^{\prime}\mathbf{,r}%
^{\prime\prime}\right)  $ in the difference $\mathbf{r}^{\prime}%
\mathbf{-r}^{\prime\prime}$ (Wigner transformation) one can express the
functional $\Phi\left[  Q\right]  $ in terms of the variables $Q_{\mathbf{p}%
}\left(  \mathbf{r}\right)  ,$ where $\mathbf{p}$ is the momentum and
$\mathbf{r}$ is the center of mass $\left(  \mathbf{r}^{\prime}+\mathbf{r}%
^{\prime\prime}\right)  /2.$ Then the free energy functional $\Phi$, Eq.
(\ref{f7}) can be written as
\begin{equation}
\Phi\lbrack Q]=\frac{i}{2}\int\mathrm{Str}\,\left[  \mathcal{H}_{J}%
(\mathbf{x})\ast Q(\mathbf{x})-i\ln Q(\mathbf{x})\right]  d\mathbf{x}
\label{f11}%
\end{equation}
where $\mathbf{x=}\left(  \mathbf{r,p}\right)  $ is the coordinate in the
phase space,
\begin{equation}
\mathcal{H}_{J}(\mathbf{x})=H_{0}\left(  \mathbf{p}\right)  +U\left(
\mathbf{r}\right)  +\frac{\omega+i\delta}{2}+iJ\left(  \mathbf{p,r}\right)
\label{f12}%
\end{equation}
is classical Hamilton function.

The product $\ast$ of two matrices $A\left(  \mathbf{x}\right)  $ and
$B\left(  \mathbf{x}\right)  $ is defined by Moyal formula%

\[
A(\mathbf{x})\ast B(\mathbf{x})=A(\mathbf{x})\mbox{e}^{\frac{i}{2}\left(
\overset{\leftarrow}{\nabla}_{\mathbf{r}}\overset{\rightarrow}{\nabla
}_{\mathbf{p}}-\overset{\rightarrow}{\nabla}_{\mathbf{r}}\overset{\leftarrow
}{\nabla}_{\mathbf{p}}\right)  }B(\mathbf{x}).
\]
The scheme of calculations using the Wigner representations and the star
product ``$\ast$'' is known as Weyl symbol calculus \cite{weyl}. This method
is convenient for quasiclassical expansions.

If the potential $U\left(  \mathbf{r}\right)  $ is smooth, one can simplify
Eq. (\ref{f11}) and come again to Eq. (\ref{sigmaaverage}). This procedure is
described in Ref. \cite{superbos}. The functional $\Phi,$ Eq. (\ref{f11}) has
a form of the Lagrangian of a non-commutative field theory \cite{weyl}. The
method suggested here can naturally be called ``superbosonization''.

At the end I have to warn that Eqs. (\ref{f5}-\ref{f7}, \ref{f11}) are not
complete yet because nothing has been said about the contour of integration
over the supermatrices $Q$. This was not very important in Refs.
(\cite{superbos,kogan1}) where a saddle point approximation was used. However,
generally, this question requires a more careful study and this is a subject
of a current work. In case if the difficulties are overcome, the
superbosonization can become very useful in the theory of random matrices for
more general models than the one described by the conventional Eq. (\ref{a1}).

\section{\textit{Summary}}

In these lectures I tried to achieve two goals:

1. The random matrix theory is to a large extent a phenomenological theory.
Therefore, it is very important to have examples when it can be obtained
starting from a microscopic model. The model of the disordered metals
considered here is the first one for which the relevance of the RMT has been
proven. The proof became possible with the help of the supermatrix non-linear
$\sigma$-model first derived for other purposes.

2. Having presented the derivation of the $\sigma$-model I demonstrated how
one obtains the Wigner-Dyson statistics from its zero-dimensional version.
However, in many situations the supersymmetry method allows to go beyond the
Wigner-Dyson model and obtain completely different results like the log-normal
distribution of the amplitudes of the wave functions of Section 2. Of course,
as soon as one has a Hamiltonian one comes to random matrices. However,
generally it is not clear how to write the distribution function for these
matrices in each particular situation and the supersymmetry method can play an
important role for investigation of microscopic models.

I wanted by no means to oppose the RMT and the supersymmetry method to each
other. They can be considered as complimentary methods, although with a
considerable overlap. We see at this school that the random matrices find more
and more applications in many fields of physics, which is an exciting
development. I believe, in many cases the supersymmetry technique can also be
useful in these new applications and one should keep in mind a possibility of
using this scheme.%

\begin{acknowledgments}
This work was supported by the Transregio 12 "Symmetries and Universality in
Mesoscopic Systems" of German Research Society".
\end{acknowledgments}



\begin{thebibliography}{10}

\bibitem{AbadiCarSecond}
M.~Abadi and L.~Cardelli.
\newblock A theory of primitive objects: Second-order systems.
\newblock {\em Science of Computer Programming}, 25(2-3):81--116, December
  1995.
\newblock A preliminary version appeared in the 1994 Proc. of European
  Symposium on Programming.

\bibitem{ArnoldGosling}
K.~Arnold and J.~Gosling.
\newblock {\em The Java Programming Language}.
\newblock Addison Wesley, 1996.


\bibitem{CardelliDot}
L.~Cardelli and X.~Leroy.
\newblock {\em Abstract types and the dot notation}, pages 479--504.
\newblock IFIP State of the Art Reports. North Holland, March 1990.
\newblock Also appeared as {SRC} {R}esearch {R}eport 56.

\bibitem{CookThesis}
W.R. Cook.
\newblock {\em A Denotational Semantics of Inheritance}.
\newblock PhD thesis, Brown University, 1989.

\bibitem{DiBlasioFisher}
P.~DiBlasio and K.~Fisher.
\newblock A concurrent object calculus.
\newblock In {\em CONCUR '96 Proc.}, pages 655--670, Pisa, 1996. Springer LNCS
  1119.


\bibitem{GoldbergR}
A.~Goldberg and D.~Robson.
\newblock {\em Smalltalk--80: The Language and its Implementation}.
\newblock Addison Wesley, 1983.

\bibitem{GunterMitch}
C.A. Gunter and J.C. Mitchell, editors.
\newblock {\em Theoretical Aspects of Object-Oriented Programming}.
\newblock MIT Press, Cambridge, MA, 1994.


\bibitem{Liskclu}
B.~Liskov et~al.
\newblock {\em {CLU} Reference Manual}.
\newblock Springer LNCS 114, Berlin, 1981.

\bibitem{LiskovS}
B.~Liskov, A.~Snyder, R.~Atkinson, and C.~Schaffert.
\newblock Abstraction mechanisms in {CLU}.
\newblock {\em Comm. ACM}, 20:564--576, 1977.

\bibitem{BMeyerEiffel}
B.~Meyer.
\newblock {\em Eiffel: The Language}.
\newblock Prentice-Hall, 1992.

\bibitem{mth90}
R.~Milner, M.~Tofte, and R.~Harper.
\newblock {\em The Definition of {Standard ML}}.
\newblock MIT Press, 1990.
\end{thebibliography}




\begin{thebibliography}{}

\bibitem[Barrett and Morton, 1984]{bm84xxx}
Barrett, J.~W. and Morton, K.~W. (1984).
\newblock Approximate symmetrization and {P}etrov-{G}alerkin methods for
  diffusion-convection problems.
\newblock {\em Comput. Methods Appl. Mech. Engrg.}, 45:97--122.

\bibitem[Chui and Wang, ]{cw:cardspline}
Chui, C.~K. and Wang, J.~Z.
\newblock A cardinal spline approach to wavelets.
\newblock {\em Proc. Amer. Math. Soc.}
\newblock to appear.

\bibitem[Daubechies, 1990]{id:signal}
Daubechies, I. (1990).
\newblock The wavelet transform, time-frequency localization and signal
  analysis.
\newblock {\em IEEE Trans. Inform. Theory}, 36:961--1005.

\bibitem[Doolan et~al., 1980]{dms80}
Doolan, E.~P., Miller, J. J.~H., and Schilders, W. H.~A. (1980).
\newblock {\em Uniform Numerical Methods for Problems with Initial and Boundary
  Layers}.
\newblock Boole Press, Dublin.

\bibitem[Garc\'{\i}a-Archilla and Mackenzie, 1991]{gm91}
Garc\'{\i}a-Archilla, B. and Mackenzie, J.~A. (1991).
\newblock Analysis of a supraconvergent cell vertex finite volume method for
  one-dimensional convection-diffusion problems.
\newblock Technical Report NA91/13, Oxford University Computing Laboratory, 11
  Keble Road, Oxford, OX1 3QD.
\newblock submitted for publication.

\bibitem[Heinrich et~al., 1977]{hhmz77}
Heinrich, J.~C., Huyakorn, P.~S., Mitchell, A.~R., and Zienkiewicz, O.~C.
  (1977).
\newblock An upwind finite element scheme for two-dimensional convective
  transport equations.
\newblock {\em Internat. J. Numer. Methods Engrg.}, 11:131--143.


\bibitem[Jameson et~al., 1981]{jst81}
Jameson, A., Schmidt, W., and Turkel, E. (1981).
\newblock Numerical solutions of the {E}uler equations by finite volume methods
  using {R}unge-{K}utta time stepping.
\newblock AIAA Paper No. 81-1259.

\bibitem[Kellogg and Tsan, 1978]{kt78}
Kellogg, R.~B. and Tsan, A. (1978).
\newblock Analysis of some difference approximations for a singular
  perturbation problem without turning points.
\newblock {\em Math. Comp.}, 32:1025--1039.

\bibitem[von Neumann, 1955]{vn:mfqm}
von Neumann, J. (1932, 1949, and 1955).
\newblock {\em Mathematical Foundations of Quantum Mechanics}.
\newblock Princeton University Press.

\end{thebibliography}


\begin{thebibliography}{9}                                                                                                %

\bibitem {wigner}E.P. Wigner, Ann. Math. \textbf{53}, 36 (1951)

\bibitem {mehta}L.M. Mehta,\textit{\ Random Matrices, }Academic Press, San
Diego (1991)

\bibitem {dyson}F.J. Dyson, J. Math. Phys. \textbf{3}, 140, 157, 166 (1962)

\bibitem {ebook}K.B. Efetov, \textit{Supersymmetry in Disorder and Chaos,
}Cambridge University Press, New York (1997)

\bibitem {verbaar}J.J.M. Verbaarschot (this volume)

\bibitem {az}A. Altland, and M.R. Zirnbauer, Phys. Rev. B \textbf{55}, 1142 (1997)

\bibitem {ge}L.P. Gorkov and G.M. Eliashberg, Zh. Eksp. Teor. Fiz. \textbf{48,
}1407 (1965) (Sov. Phys. JETP \textbf{21}, 940 (1965))

\bibitem {agd}A.A. Abrikosov, L.P. Gorkov, and I.E. Dzyaloshinskii,
\textit{Methods of Quantum Field Theory in Statistical Physics, }Prentice
Hall, New York (1963)\textit{\ }

\bibitem {kubo1}R. Kubo, J. Phys. Soc. Jpn. \textbf{17}, 975 (1962)

\bibitem {kubo2}R. Kubo, in \textit{Polarization, Matiere et Rayonnement,
}livre jubile en l'honneur du Professeur A. Kastler, Presses Universitaires de
France, Paris, p.325 (1969)

\bibitem {aalr}E. Abrahams, P.W. Anderson, D.C. Licciardello, and T.V.
Ramakrishnan, Phys. Rev. Lett. \textbf{42}, 673 (1979)

\bibitem {glk}L.P. Gorkov, A.I. Larkin, and D.E. Khmelnitskii, Pis'ma Zh.
Eksp. Teor. Fiz. \textbf{30, }248 (1979) (Sov. Phys. JETP Lett. \textbf{30},
228 (1979))

\bibitem {efetov1}K.B. Efetov, Zh. Eksp. Teor. Fiz. \textbf{82}, 872 (1982)
(Sov. Phys. JETP \textbf{55, }514 (1982); J. Phys. C \textbf{15}, L909 (1982);
Zh. Eksp. Teor. Fiz. \textbf{83}, 833 (Sov. Phys. JETP \textbf{56}, 467 (1982)

\bibitem {ereview}K.B. Efetov, Adv. Phys. \textbf{32}, 53 (1983)

\bibitem {vwz}J.J.M. Verbaarschot, H.A. Weidenm\"{u}ller, and M.R. Zirnbauer,
Phys. Rep. \textbf{129}, 367 (1985)

\bibitem {bgs}O. Bohigas, M.J. Gianonni and C. Schmidt, Phys. Rev. Lett.
\textbf{52}, 1 (1984); J. Physique Lett. \textbf{45}, L1615 (1984)

\bibitem {suptrace}\textit{Supersymmetry and Trace Formula, (}edited
by\textit{\ }J.P. Keating, D.E. Khmelnitskii, and I.V. Lerner), NATO\ ASI
Series B: Physics Vol. \textbf{370}, Kluwer Academic, New York (1999)

\bibitem {wegner}F. Wegner, Z. Phys. B \textbf{35}, 207(1979)

\bibitem {elk}K.B. Efetov, A.I. Larkin, D.E. Khemelnitskii, Zh. Eksp. Teor.
Fiz. \textbf{79}, 1120 (1980) (Sov. Phys. JETP, \textbf{52}, 568 (1980))

\bibitem {sw}L. Sch\"{a}fer and F. Wegner, Z. Phys. B \textbf{38}, 113 (1980)

\bibitem {km}A. Kamenev and M. Mezard, J. Phys. A: Math. Gen. \textbf{32},
4373 (1999)

\bibitem {yl}I.V. Yurkevich and I.V. Lerner, Phys. Rev. B \textbf{60}, 3955 (1999)

\bibitem {kanzieper}E. Kanzieper, Phys. Rev. Lett. \textbf{89}, 250201 (2002)

\bibitem {finkelstein}A.M. Finkel'stein, Zh. Eksp. Teor. Fiz. \textbf{84}, 168
(1983) (Sov. Phys. JETP \textbf{57}, 97 (1983))

\bibitem {fin1}A.M. Finkel'stein, in \textit{Electron Liquid in Disordered
Conductors, }edited by I.M. Khalatnikov, Soviet Scientific Reviews, Vol. 14
(Harwood, London, 1990)

\bibitem {schwiete}G. Schwiete and K.B. Efetov, Phys. Rev. B, 2005 (in press), cond-mat/0409546

\bibitem {fyodorov}Y.V. Fyodorov, H.J. Sommers, J. Math. Phys. \textbf{38},
1918 (1997)

\bibitem {gmw}T. Guhr, A. Muller-Groeling, H.A.Weidenmuller, Phys. Rep.
\textbf{299}, 190 (1998)

\bibitem {mirlin}A.D. Mirlin, Phys. Rep. \textbf{326}, 259 (2000)

\bibitem {ast}A. Altland, B.D. Simons, D. Taras-Semchuk, Adv. Phys.
\textbf{49}, 321 (2000)

\bibitem {asz}A. Altland, B.D. Simons, M.R. Zirnbauer, Phys. Rep.
\textbf{359}, 283 (2002)

\bibitem {ps}G. Parisi and N. Sourlas, Pys. Rev. Lett. \textbf{43}, 744 (1979)

\bibitem {berezin}Dokl. Akad. Nauk SSSR, \textbf{137}, 31 (1961);
\textit{Metod vtorichnogo kvantovaniya,} (\textit{The method of Second
Quantization), }Nauka (1965)\textit{; Introduction to Superanalysis, }MPAP,
Vol. 9 D. Reidel, Dordrecht (1987)

\bibitem {altshk}B.L. Altshuler and B.I. Shklovskii Zh. Eksp. Teor. Fiz.
\textbf{91}, 220 (1986) (Sov. Phys. JETP \textbf{64}, 127 (1986)

\bibitem {el}K.B. Efetov and A.I. Larkin, Zh. Eksp. Teor. Fiz. \textbf{85},
764 (1983) (Sov. Phys. JETP, \textbf{58}, 444 (1983)

\bibitem {fm}Y.V. Fyodorov and A.D. Mirlin, Phys. Rev. Lett. \textbf{67}, 2405 (1991)

\bibitem {eb}{K.B. Efetov, }Zh. Eksp. Teor. Fiz. \textbf{88}, 1032-1052
(1985); (Sov. Phys. JETP. \textbf{61}, 606-617, 1985)\newline

\bibitem {mf}A.D. Mirlin and Y.V. Fyodorov, J. Phys. A \textbf{24}, 2273 (1991)

\bibitem {mp}M. Mezard and G. Parisi, J. Phys. (Paris) \textbf{47}, 1285
(1985); Europhys. Lett. \textbf{3}, 1067 (1987)

\bibitem {fe}V.I. Falko and K.B. Efetov, Europhys. Lett. \textbf{32}, 627
(1995); Phys. Rev. B\textbf{52}, 17413 (1995)

\bibitem {porter}C.E. Porter, \textit{Statistical Theories of
Spectra:\ Fluctuations, }Academic Press, New York (1965)

\bibitem {brody}T.A. Brody, J. Flores, J.B. French, P.A. Mello, A. Panday, and
S.S.M. Wong, Rev. Mod. Phys. \textbf{53}, 385 (1981)

\bibitem {polyakov}A.M. Polyakov, Phys. Lett. \textbf{103B }207, 211 (1981)

\bibitem {seiberg}N. Seiberg, Prog. Theor. Phys. \textbf{102 }Supl., 319 (1990)

\bibitem {tsvelik}A. Tsvelik, C. Mudry, and Y. Kogan, Phys. Rev. Lett,
\textbf{77, }707 (1996)

\bibitem {mkh}B.A. Muzykantskii and D.E. Khmelnitskii, Phys. Rev. B
\textbf{51}, 5480 (1995)

\bibitem {fmir}Y.V. Fyodorov and A.D. Mirlin, JETP Lett. \textbf{60}, 790 (1994)

\bibitem {akl}B.L. Altshuler, V.E. Kravtsov, and I.V. Lerner, Pis'ma Zh. Eksp.
Teor. Fiz. \textbf{43}, 342 (1986) (Sov. Phys. JETP Lett. \textbf{43, }441
(1986)); Zh. Eksp. Teor. Fiz. \textbf{91}, 2276 (1986) (Sov. Phys. JETP
\textbf{64}, 1352 (1986)); Phys. Lett. A \textbf{134, }488 (1989)

\bibitem {been}C.W.J. Beenakker, Rev. Mod. Phys. \textbf{69}, 733 (1997)

\bibitem {eilen}G. Eilenberger, Z. Phys. \textbf{214}, 195 (1968)

\bibitem {mk}B. A. Muzykantskii and D. E. Khmelnitskii, Pis'ma Zh. Eksp. Theo.
Fiz. \textbf{62} 68 (1995) [JETP Lett. \textbf{62} 76 (1995)]; (also see in
Ref. \cite{suptrace}).

\bibitem {kogan}K. B. Efetov, V. R. Kogan, Phys. Rev. B \textbf{67}, 245312 (2003)

\bibitem {aos}A. Altland, C.R. Offer, and B.D. Simons in \cite{suptrace}.

\bibitem {al}I.L. Aleiner and A.I. Larkin, Phys. Rev. B\textbf{54}, 14423
(1996); Phys. Rev. E \textbf{55} (2): R1243 (1997)

\bibitem {superbos}K. B. Efetov, G. Schwiete, K. Takahashi, Phys. Rev. Lett.
\textbf{92}, 026807 (2004)

\bibitem {boson}P.B. Wiegmann, Nucl. Phys. B \textbf{323}, 311 (1989); S.R.
Das, A. Dhar, G. Mandal, and S.R. Wadia, Mod. Phys. Lett. A \textbf{7}, 71
(1992); A. Dhar, G. Mandal, and S.R. Wadia, \textit{ibid.} \textbf{7}, 3129
(1992); D.V. Khveshchenko, Phys. Rev. B \textbf{49}, 16893 (1994);
\textit{ibid.} \textbf{52}, 4833 (1995).

\bibitem {kogan1}K. B. Efetov, V. R. Kogan, Rev. B \textbf{70}, 195326 (2004)

\bibitem {weyl}S.W. Mc Donald, Phys. Rep. \textbf{158}, 337 (1988)
\end{thebibliography}
\end{document}